\documentclass[twocolumn,PRD,showpacs,nofootinbib,amsmath,amssymb,superscriptaddress]{revtex4-1}

\pdfoutput=1

\usepackage{graphicx}
\usepackage{dcolumn}
\usepackage{bm}
\usepackage{hyperref}
\usepackage{color}
\usepackage{rotate}
\usepackage{fancyhdr}
\usepackage{cancel}
\usepackage{multirow}
\usepackage{ulem}

\def\be{\begin{equation}}
\def\ee{\end{equation}}
\def\bea{\begin{eqnarray}}
\def\eea{\end{eqnarray}}

\newcommand{\beq}{\begin{equation}}
\newcommand{\eeq}{\end{equation}}

\definecolor{darkred}{RGB}{175,0,0}
\definecolor{darkblue}{RGB}{14,0,185}

\usepackage{aas_macros,braket}
\begin{document}
\title{Robustness of baryon acoustic oscillation constraints for early-Universe modifications to $\Lambda$CDM}
\author{Jos\'e Luis Bernal}
\affiliation{Department of Physics and Astronomy, Johns Hopkins University, 3400 North Charles Street, Baltimore, Maryland 21218, USA}
\author{Tristan L. Smith}
\affiliation{Department of Physics and Astronomy, Swarthmore College, 500 College Ave., Swarthmore, PA 19081, USA}
\author{Kimberly K. Boddy}
\affiliation{Theory Group, Department of Physics, The University of Texas at Austin, Austin, TX 78712, USA}
\author{Marc Kamionkowski}
\affiliation{Department of Physics and Astronomy, Johns Hopkins University, 3400 North Charles Street, Baltimore, Maryland 21218, USA}

\begin{abstract}
Baryon acoustic oscillations (BAO) provide a robust standard ruler, and  can be used to constrain the expansion history of the Universe at low redshift. Standard BAO analyses return a model-independent measurement of the expansion rate and the comoving angular diameter distance as function of redshift, normalized by the sound horizon at radiation drag. However, this methodology relies on anisotropic distance distortions of a fixed, pre-computed template (obtained in a given fiducial cosmology) in order to fit the observations. Therefore, it may be possible that extensions to the consensus $\Lambda$CDM add contributions to the BAO feature that cannot be captured by the template fitting. We perform mock BAO fits to power spectra computed assuming cosmological models which modify the growth of perturbations prior to recombination in order to test the robustness of the standard BAO analysis. We find no significant bias in the BAO analysis for the models under study ($\Lambda$CDM with a free effective number of relativistic species, early dark energy, and a model with interactions between neutrinos and a fraction of the dark matter), even for cases which do not provide a good fit to \textit{Planck} measurements of the cosmic microwave background power spectra. This result supports the use of the standard BAO analysis  and its measurements to perform cosmological parameter inference and to constrain exotic models. In addition, we provide a methodology to reproduce our study for different models and surveys, as well as discuss different options to handle eventual biases in the BAO measurements.  
\end{abstract}

\maketitle

\section{Introduction}\label{sec:Intro}
Baryon acoustic oscillations (BAO) appear due to the primordial sound waves propagating in the tightly coupled photon-baryon plasma in the early Universe until recombination (e.g.,~\cite{Peebles_perturbations,Sunyaev_fluctuations}). After recombination, when the interaction rate between photons and baryons becomes inefficient due to Hubble expansion, the acoustic waves stop propagating and the BAO are imprinted in the baryon distribution. Therefore, BAO are present in the temperature anisotropies of the cosmic microwave background (CMB), which allows a very precise inference of cosmological parameters (see e.g.,~\cite{Planck18_pars}). In addition, their features also appear in the matter (and subsequently the galaxy) distribution at low redshift, although with lower significance. First detected in the galaxy power spectrum around fifteen years ago~\cite{firstBAO_SDSS,firstBAO_2dF},  
  BAO have been robustly measured in galaxy, quasar, and Lyman-$\alpha$ density distributions reaching percent-level precision (see e.g,~\cite{Alam_bossdr12,GilMarin_QSOeBOSS,eBOSS_Lyalpha_auto,eBOSS_Lyalpha_cross}).

The BAO features are characterized by a physical scale: the sound horizon at radiation drag, $r_{\rm d}$. 
With this `standard ruler', it is possible to robustly map the expansion history of the Universe up to the redshift of measurement~\cite{Eisenstein_02ASP,Blake_standardruler,Seo_BAODE}. Given that BAO measurements depend on both $r_{\rm d}$ and the expansion history, they are sensitive to early and late time physics in clean, distinct ways. 

The BAO and type Ia supernovae (SNeIa)~\cite{Betoule14_jla,Scolnic_pantheon} have become the main complement to CMB observations to constrain deviations from the $\Lambda$CDM consensus model at low redshift. Moreover, since BAO measurements form a standard ruler which connects the direct and inverse distance ladders~\cite{Cuesta:2014asa},  they have been key in addressing the Hubble constant ($H_0$) tension~\cite{RiessH0_19, H0_holicow_19, Planck18_pars, Addison_BAOBBN, Schoneberg_BAOBBN, Verde_H0summary}. As pointed out in Ref.~\cite{BernalH0} (and confirmed by independent analyses~\cite{Poulin_H0,Aylor_sounds}), the $H_0$ tension can be reframed as a mismatch between the anchors of the direct and the inverse cosmic distance ladders ($H_0$ and $r_{\rm d}$, respectively), with existing data constraining deviations of the evolution of the expansion history at low redshift. This has led authors of Ref.~\cite{Know_H0hunter} to theorize that  any extension to $\Lambda$CDM should modify pre-recombination physics rather than the low-redshift Universe if it is to alleviate the $H_0$ tension. Some examples include Refs.~\cite{Poulin_EDE,Smith_EDE,Kreisch_selfnu,Ghosh_DNIH0,Lin_acousticDM,Agrawal_R&R,Berghaus_H0friction,Zumalacarregui_MGH0}.

Standard BAO analyses rely on templates of the summary statistic under study computed assuming a fiducial cosmology (normally $\Lambda$CDM). They consist of determining the anisotropic rescaling of the distance coordinates that the template needs in order to fit observations, marginalizing over a plethora of nuisance parameters to account for other deviations with respect to the template.  This approach allows for a tomographic, model-independent determination of the expansion history of the Universe that can be used to constrain different cosmological models. 

 This procedure has been proven to be extremely robust and flexible for models predicting different expansion rates at late times (see e.g.,~\cite{Xu_anisoBAO,Aubourg15,VargasMagana_BAOtheosyst, Carter_BAOtest}), and it successfully models  changes in $r_{\rm d}$ due to early-time modifications of the cosmological model. However, there may be other contributions to the BAO feature that are not captured by these rescaling and nuisance parameters, such as phase shifts (which can be scale dependent) or a different scale dependence of the amplitude of the oscillations. These additional contributions may bias  standard BAO analyses (in the case they are not properly modeled in the template), but they can also be targeted to  constrain beyond-$\Lambda$CDM physics with BAO  (as shown in Refs.~\cite{Baumann_NeffBAO,Baumann_NeffBAO_fisher,Baumann_NeffBAO_measure}  for the case of extra relativistic species $N_{\rm eff}$, and in Ref.~\cite{Beutler_PrimBAO} for the case of oscillations in the primordial power spectrum). Given the key role of BAO measurements in constraining extensions to $\Lambda$CDM,  it is essential to ascertain whether cosmological-parameter inference from BAO analyses is free of unaccounted systematic errors, especially when the requirements on the robustness of the analyses will tighten for next-generation BAO measurements. 

 In this work we explore the impact that modifications of the BAO feature, arising from changes in the growth of the matter density perturbations, may have in a BAO analysis. Concretely, we consider the possibility that standard template-fitting BAO analyses are not flexible enough to characterize these modifications. If this were the case, the reported BAO measurements obtained using a $\Lambda$CDM template might be biased if the true cosmology corresponds to these models; moreover, a bias could render BAO measurements unsuitable for constraining extended models.
 
 This study is therefore timely and needed in order to ensure the robustness of using BAO measurements to constrain models beyond $\Lambda$CDM.  We focus on $\Lambda$CDM, a model with a different number of relativistic species ($\Lambda$CDM$+N_{\rm eff}$), and two models that claim to ease or resolve the Hubble tension: early dark energy (EDE)~\cite{Poulin_EDE,Smith_EDE} (although see also Refs.~\cite{Hill_EDE,Ivanov_EDEEFT,D'Amico_EFTEDE, SmithEDELSS,PoulinEDEWL}) and dark neutrino interactions (DNI)~\cite{Ghosh_DNIH0,Ghosh_DNIGW}.  We perform BAO analyses using mock power spectrum measurements computed under these models, for both `good' and `bad' fits to CMB observations. We use the term `bad fits'  to refer to sets of cosmological parameters for which one of the parameters has been chosen to have a fixed value at least $\gtrsim 3\sigma$ away from its best fit to \textit{Planck}, while the rest of the parameters are selected to maximize the posterior fulfilling such requirement. We also include an extreme case for DNI, the model adding the largest difference in the BAO feature with respect to the $\Lambda$CDM prediction. We find that the standard BAO analysis returns results without significant biases, even when considering the `bad fits' mock power spectra: the biases in the BAO rescaling parameters are $<0.2\sigma$ except for the extreme case of DNI, for which we find a bias $\sim  1\sigma$). 
 These results confirm the robustness of BAO analyses and their appropriate use for cosmological parameter inference.  An alternative, general approach to estimate the bias on parameter inference can be found in e.g., Ref.~\cite{Bernal_Fisher}. 
   
   We describe in detail our methodology for future application to other models to check whether the standard BAO analysis is unbiased. In case a bias were to be found, we discuss different courses of action to address the situation. Nonetheless, we also provide approximate arguments for extrapolating our results to other cosmological models. 
 
Although this work focuses on the BAO feature, there is also cosmological information contained in the broadband of the observed clustering. Redshift-space distortions at linear scales  are mostly sensitive to the product $f\sigma_8$, where $f$ is the linear growth rate and $\sigma_8$ is the root mean square of the matter fluctuations within $8\, {\rm Mpc}/h$ (where $h = H_0/\left(100\, {\rm km\,s^{-1}Mpc^{-1} }\right)$). The degeneracy between $f$ and $\sigma_8$ can be broken by e.g., using higher-order statistics, such as the three-point correlation function~\cite{Slepian_3PCFmodel} or the bispectrum~\cite{GilMarin_RSDBispectrumDR12,Gualdi_bispectrummultipoles},  by exploiting the clustering at smaller scales using perturbation theory (see e.g.,~\cite{Desjacques_bias}) or by using phase correlations~\cite{Wolstenhulme_phases, Byun_phase}.  Additional physics related to large modifications of the scale dependence of the clustering can be strongly constrained, such as suppression at small scales (see e.g.,~\cite{Drlica-Wagner_LSSTDM, Bullock_smallscalesreview, Gluscevic_DMintWP, Nadler_DMMW,  Abazajian_sterilenureview}  and references therein) or primordial non-Gaussianities (see e.g.,~\cite{Dalal_png07,Matarrese_png08,Raccanelli:fNL,Bernal_EMU}). Parameters driving features with weaker scale dependence  can also be constrained, but normally  external data or priors are required for these constraints to be competitive. 
 
 This paper is structured as follows: we discuss the origin and nature of the cosmological information encoded in the BAO measurements in Section~\ref{sec:info}; review the standard BAO analysis of the power spectrum in Section~\ref{sec:analysis}; introduce the cosmological models considered, discuss the variations in the predicted BAO feature, and estimate the bias in the BAO fit in Section~\ref{sec:bias_estimation}; and discuss the results and conclude in Sections~\ref{sec:discussion} and \ref{sec:conclusions}, respectively. Afterwards, we discuss the methodologies used to extract and isolate the BAO feature in Appendix~\ref{sec:Olin_templ};  detail how we compute the mock measured power spectrum in Appendix~\ref{sec:mockdata}; and  illustrate the  connection between the CMB power spectra and the BAO feature in the matter power spectrum at late times with a toy model in Appendix~\ref{sec:CMBtoBAO}.

\section{BAO Cosmology}\label{sec:info}
In this section, we describe the cosmological information contained in the BAO feature of the low-redshift  clustering of tracers of matter density perturbations. 
 The analysis of large-scale structure clustering allows for precise determination of cosmological parameters. By separating the clustering at different scales into a smooth, featureless broadband and the BAO feature, it is possible to extract robust cosmological information encoded in the BAO scale.

\subsection{Alcock-Paczynski effect and isotropic dilation}
Observations measure the positions of different tracers of matter in terms of redshifts and angular positions on the sky, which must then be transformed to obtain three-dimensional clustering summary statistics (e.g., the correlation function or power spectrum) as a function of spatial distances or the corresponding Fourier mode wave numbers. Given an angular separation $\theta$ and a small redshift separation $\delta z$, the spatial comoving distance in the transverse direction and along the line of sight are 
\begin{equation}
r_\perp =D_M(z)\theta,\qquad\qquad r_\parallel = \frac{c\delta z}{H(z)},
\label{eq:z_to_distances}
\end{equation} 
respectively, where $D_M$ is the comoving angular diameter distance, $H$ is the Hubble expansion rate, and $c$ is the speed of light. Eq.~\eqref{eq:z_to_distances} can be adapted to proper distances substituting $D_M$ by the proper angular diameter distance $D_A=D_M/(1+z)$.  
There are three main effects that alter these components of the observed distances: redshift-space distortions~\cite{Hamilton_RSDreview}, the Alcock-Paczynski effect~\cite{Alcock_paczynski} and the isotropic dilation.
Redshift-space distortions are a physical modification to $r_\parallel$, due to the peculiar velocities of galaxies changing the redshift of observed sources along the line of sight (hence changing their position in redshift space with respect to the real space). These distortions introduce anisotropies in the observed clustering, which is isotropic a priori.

Assuming a background expansion history (obtained from a \textit{fiducial} cosmology)  that differs from that of the \textit{true} expansion rate of the Universe causes an artificial distance distortion.  The fiducial cosmology is used to compute $D_M$ and $H$ in Eq.~\eqref{eq:z_to_distances}; therefore, the recovered $r_\perp$ and $r_\parallel$ differ from the true distances.  This distortion  affects $r_\perp$ and $r_\parallel$ in different ways, so it is possible to decompose it into an isotropic and an anistropic component: the isotropic dilation and the Alcock-Paczynski effect.

The Alcock-Paczynski effect and the isotropic dilation can be modeled by rescaling factors, obtained when comparing the \textit{observed} distances, which assume the fiducial cosmology, and \textit{true} distances: $r_{\perp,\parallel}^{\rm true} = r^{\rm obs}_{\perp,\parallel}q_{\perp,\parallel}$ (or $k^{\rm true}_{\perp,\parallel}=k^{\rm obs}_{\perp,\parallel}/q_{\perp,\parallel}$ in Fourier space). Using Eq.~\eqref{eq:z_to_distances}, the rescaling parameters are
\begin{equation}
q_\perp = \frac{D_M(z)}{\left( D_M(z)\right)^{\rm fid1}},\qquad q_\parallel = \frac{\left(H(z)\right)^{\rm fid1}}{H(z)},
\label{eq:scaling_q}
\end{equation}
where superscript `fid1' denotes the assumed fiducial cosmology used in Eq.~\eqref{eq:z_to_distances}. Using $q_\perp$ and $q_\parallel$, the isotropic dilation corresponds to $\left(q_\perp^2q_\parallel\right)^{1/3}$, and the Alcock-Paczinski effect is given by the ratio of $q_\perp$ and $q_\parallel$. 

\subsection{BAO scale}
The Alcock-Paczysnki effect and the isotropic dilation are always present\footnote{The only case in which the rescaling of distances in Eq.~\eqref{eq:scaling_q} is not  explicitly present in the analysis is if the clustering statistics used are functions of redshifts and angles (see e.g., Ref.~\cite{Bertacca_xi3dGR}). In this case, the  rescaling is implicitly embedded in the modeling of the clustering statistics.} in the measurement of the clustering statistics: it is inherent to any measurement that depends on distance scales.  Nonetheless, the BAO feature is clearly distinguishable against the broadband of the summary statistic; it manifests as oscillations in Fourier space or a peak in configuration space, and large-scale clustering measurements have well-determined its location.

To extract the BAO scale from the observed target summary statistic, standard BAO analyses employ a pre-computed template of the target summary  statistic generated assuming a given cosmology. This cosmology does not need to be the same as that  used in Eqs.~\eqref{eq:z_to_distances} and~\eqref{eq:scaling_q}. As we discuss in further detail in Section~\ref{sec:analysis}, using the template allows for the extraction of $r_{\rm d}$, which is the only characteristic scale of matter clustering at low redshifts (at larger scales than those corresponding to matter-radiation equality). The BAO scale of the template might not match the true BAO scale; therefore, a correction on $r_{\rm d}$ must be included when rescaling distances in order to fit the observed BAO feature with the template. The correction is isotropic, and the rescaling of distances becomes $r_{\perp,\parallel}^{\rm th} = r^{\rm obs}_{\perp,\parallel}\alpha_{\perp,\parallel}$ (or $k^{\rm th}_{\perp,\parallel}=k^{\rm obs}_{\perp,\parallel}/\alpha_{\perp,\parallel}$), where
\begin{equation}
\alpha_\perp = q_\perp\frac{\left(r_{\rm d}\right)^{\rm fid2}}{r_{\rm d}},\qquad \alpha_\parallel = q_\parallel\frac{\left(r_{\rm d}\right)^{\rm fid2}}{r_{\rm d}}
\label{eq:scaling_2cosmo}
\end{equation}
provide a mapping between the observed distances (or wave numbers) and those which enter our theoretical modeling, denoted by `th'.\footnote{There are alternative parametrizations of these rescalings (or those in Eq.~\eqref{eq:scaling_q}), obtained through combinations of $\alpha_\perp$ and $\alpha_\parallel$. Some examples focus on the isotropic and anisotropic distortions $(\alpha,\epsilon)$ or on the monopole and the $\mu^2$ moment of the two-point statistics $(\alpha_0,\alpha_2)$ (see e.g., Refs.~\cite{Xu_anisoBAO} and~\cite{GilMarin_BAODR12}, respectively).} The superscript `fid2' in Eq.~\eqref{eq:scaling_2cosmo} refers to the fiducial cosmology used to generate the template for the BAO analysis.  It is important to notice that the rescaling of $r_{\rm d}$ in Eq.~\eqref{eq:scaling_2cosmo} is not related to the Alcock-Paczynski effect or the isotropic dilation.  Hence, the rescaling between observed distances and those entering our theoretical model introduced in Eq.~\eqref{eq:scaling_2cosmo} is the combination of two non-physical effects: the  redshift-distance transformation and the ratio between the fiducial (for the fixed template) and true $r_{\rm d}$ values. 

Hereinafter we assume that the fiducial cosmology used to convert redshifts into distances in Eq.~\eqref{eq:z_to_distances} is the same as the one used to compute the template of the clustering statistic, as is commonly the case in BAO studies. Under this assumption, the rescaling parameters become
\begin{equation}
\begin{split}
\alpha_\perp  = \frac{D_M(z)/ r_{\rm d}}{\left(D_M(z)/r_{\rm d}\right)^{\rm fid}},\qquad
 \alpha_\parallel  = \frac{\left(H(z)r_{\rm d}\right)^{\rm fid}}{H(z)r_{\rm d}},
\end{split} 
\label{eq:scaling}
\end{equation}
where `fid' corresponds to the fiducial cosmology that has been used to both translate redshifts into distances and compute the fixed template. 

\subsection{Cosmological information in BAO}
 In order to avoid biasing the information obtained from the BAO feature, the shape and amplitude of the broadband are marginalized over with the introduction of nuisance parameters. After marginalization, the only remaining cosmological information in the clustering statistics is related to the BAO location and anisotropy, which is mostly encoded in the rescaling parameters.   This is an entirely geometric fit to the observations; hence, it has the potential to be performed without being limited to any cosmological model without loss of generality. Specifically, the rescaling parameters are the fit parameters, and the resulting constraints are traditionally used in global analyses to infer cosmological parameters of any cosmological model.

As evident from Eq.~\eqref{eq:scaling_q}, the only cosmological information the Alcock-Paczynski effect and the isotropic dilation are sensitive to is the late-time expansion rate. 
By utilizing a fixed template in the analysis, BAO measurements are also sensitive to pre-recombination physics through $r_{\rm d}$ (Eq.~\eqref{eq:scaling}) in an agnostic and independent way, incorporating information about both the expansion rate and the growth of matter perturbations. While the isotropic dilation is completely degenerate with $r_{\rm d}$, the Alcock-Paczynski effect (modeled by the ratio of $\alpha_\perp$ and $\alpha_\parallel$ when a fixed template is used) is independent of the BAO scale. Since $\alpha_\perp/\alpha_\parallel=D_MH/\left(D_MH\right)^{\rm fid}$ does not depend on $H_0$, the Alcock-Paczynski effect constrains the unnormalized expansion history of the Universe, independently of the BAO scale. 

Accessing early-time information about $r_{\rm d}$ enables the construction of the inverse distance ladder~\cite{Cuesta:2014asa} and allows for an independent constraint on the product $r_{\rm d}H_0$~\cite{StandardQuantities}, which is what makes the BAO measurements key for the $H_0$ tension~\cite{BernalH0}.
While having sensitivity to $r_{\rm d}$ is beneficial, the fixed template may reduce the flexibility of the analysis and, therefore, the applicability of the constraints on $\alpha_{\perp,\parallel}$ to cosmological models that deviate from the fiducial cosmology in the early Universe.
We explore this possibility in Section~\ref{sec:bias_alphas}.

Finally, there is also cosmological information contained not only in the location of the BAO feature, but also in its shape and amplitude (see e.g.,~\cite{Baumann_NeffBAO, Beutler_PrimBAO}).  However, given that the significance of the BAO feature with respect to the broadband in the density distribution at low redshift is significantly smaller than in the CMB power spectra, the constraining power of this information is limited with respect to the CMB constraints.  Nonetheless, for some cases, external priors on the rescaling parameters can be used to practically fix their values and focus only on the shape of the BAO. This approach has provided promising results to constrain $N_{\rm eff}$~\cite{Baumann_NeffBAO_fisher,Baumann_NeffBAO_measure} and oscillations in the primordial power spectrum~\cite{Beutler_PrimBAO}, and it offers an alternative (but not independent) approach to look for physics beyond $\Lambda$CDM, complementary to direct CMB constraints.

\section{BAO analysis}\label{sec:analysis}
In this section, we introduce the methodology that we use to extract the BAO scale in this work. We follow the standard analysis used to obtain BAO measurements from galaxy surveys, using the galaxy power spectrum in Fourier space. A similar analysis can be performed in configuration space using the correlation function (see e.g.,~\cite{Anderson_BOSSDR9} for the computation of the templates).

\subsection{Power spectrum template and methodology}
The power spectrum $P(\bm{k})$ and the correlation function $\xi(\bm{s})$ are equivalent estimators for two-point clustering statistics  in  Fourier and  configuration space, respectively, where $\bm{s}$ is the redshift space distance and $\bm{k}$ is the associated wave number. The Legendre multipoles of the power spectrum are given by
\begin{equation}
P_\ell(k)=\frac{2\ell+1}{2}\int_{-1}^1{\rm d}\mu P(k,\mu)\mathcal{L}_\ell(\mu),
\end{equation}
where $k\equiv \lvert \bm{k}\lvert$ is the module of the wave number vector and $\mu$ is the cosine of the angle between the wave number vector and the line of sight. The Legendre multipoles of the correlation function, $\xi_\ell$, are defined in an analogous way, and related with $P_\ell$  by the Fourier transform via
\begin{equation}
\xi_{\ell}(s) = i^\ell\int\frac{k^3{\rm d}\log k}{2\pi^2}P_{\ell}(k)j_\ell(ks),
\label{eq:pktoxi}
\end{equation} 
where $s\equiv \lvert \bm{s}\lvert$ is the module of the redshift space distance and $j_\ell$ is the $\ell$-th order spherical Bessel function. Note that  Eq.~\eqref{eq:pktoxi} equally holds for real space distances and wave numbers.  In this section,  we do not explicitly include the dependence on redshift, present in practically all quantities, for the sake of brevity and readability; we do, however, show the dependence on $k$ and $\mu$, for clarity. 

The standard BAO analysis is based on fitting a template (pre-computed under a fiducial cosmology) to the observations. This template is built in such a way that the BAO feature is identifiable and isolated.  In order to isolate the BAO feature, the linear matter power spectrum $P_{\rm m}$  is decomposed into a smooth component $P_{\rm m,sm}$ (i.e., the broadband, with no contribution from the BAO) and an oscillatory contribution $O_{\rm lin}$. In this way, the total matter power spectrum is given by $P_{\rm m}(k) = P_{\rm m,sm}(k)O_{\rm lin}(k)$. There are different ways to extract $P_{\rm m,sm}$ from $P_{\rm m}$, and we describe the methodology used here in Appendix~\ref{sec:Olin_templ}.\footnote{Our methodology is a generalization of the approach introduced in Ref.~\cite{Kirkby_BAOtemplate}, accounting for the fact that $r_{\rm d}$ can have (very) different values depending on the cosmology assumed.} 
 
  The galaxy bias $b_{\rm g}$ and a factor encoding the effect of redshift-space distortions  $F_{\rm RSD}$ can be applied to $P_{\rm m,sm}$ in order to obtain the anisotropic, smoothed galaxy power spectrum in redshift space
  \begin{equation}
P_{\rm sm}(k,\mu) = BF_{\rm RSD}^2(k,\mu)P_{\rm m,sm}(k),
\label{eq:Psmk}
\end{equation}
where  $B$ is a constant absorbing  $b_{\rm g}$ and potential variations on the amplitude of $P_{\rm m,sm}$, and 
\begin{equation}
F_{\rm RSD}(k,\mu) = \left(1+ \beta\mu^2R\right)\frac{1}{1+0.5\left(k\mu\sigma_{\rm FoG}\right)^2},
\label{eq:RSD}
\end{equation}
where $\beta = f/b_g$, and the fingers of God small-scale suppression is driven by the parameter  $\sigma_{\rm FoG}$, whose value is related to the halo velocity dispersion.\footnote{The damping due to the fingers of God can also be modeled with a Gaussian function, providing similar results without losing flexibility in the fit to the observations.}  

 The actual amplitude of the BAO feature is reduced with respect to the linear prediction due to non-linear collapse. In addition, non-linear clustering also introduces a sub-percent shift in the BAO scale. However, these effects  can be partially reverted using density field reconstruction~\cite{Eisenstein_reconstruction,Padmanabhan_recons}. The effect of wrong assumptions regarding the fiducial cosmology and galaxy bias on density field reconstruction was studied in Ref.~\cite{Sherwin_wrongrecon}, where a negligible shift in the BAO peak location was found. However, it was also found that these small  changes in the shape of the BAO feature and in the amplitude of the quadrupole of the power spectrum might introduce small biases in the BAO fit for future surveys. The factor $R$ in Eq.~\eqref{eq:RSD} models the partial removal of redshift-space distortions produced by the density field reconstruction and takes the following values: $R=1$ before reconstruction and $R=1-\exp\left[-\left(k\Sigma_{\rm recon}\right)^2/2\right]$ after reconstruction.\footnote{The specific functional form of $R$ after reconstruction depends on the reconstruction formalism used~\cite{Seo_BAOrecon}. The one used in this work corresponds to the `Rec-Iso' convention in Ref.~\cite{Seo_BAOrecon}, which accounts for the removal of redshift-space distortions during the process of reconstruction. Nonetheless, the choice of reconstruction formalism does not affect our results, since we use the same approach for the analysis and for the computation of the mock power spectra that we analyze.}  On the other hand, the non-linear damping of the BAO is modeled with an exponential suppression applied to $O_{\rm lin}$~\cite{Eisenstein_BAOrobust,Crocce_BAOdamp,Seo_nonlinearBAO,Seo_nonlinearBAO10}. The damping affects the transverse and line-of-sight directions differently; hence, we introduce two separate scales $\Sigma_\perp$ and $\Sigma_\parallel$, respectively.  
 
The final anisotropic galaxy power spectrum, accounting for the effect of non-linearities on the BAO features and eventual density field reconstruction, can be expressed as
\begin{equation}
\begin{split}
 P(k,\mu)& = P_{\rm sm}(k,\mu)\, \times\\
\times &\left[1+\left(O_{\rm lin}(k)-1\right)e^{ -\frac{k^2}{2}\left\lbrace\mu^2\Sigma_\parallel^2+\left(1-\mu^2\right)\Sigma_\perp^2\right\rbrace}\right] + \\
+ & P_{\rm shot}\,, 
\end{split}
\label{eq:Pkmu}
\end{equation}
where $P_{\rm shot}=n_{\rm g}^{-1}$ (where $n_{\rm g}$ is the mean  comoving  number density of galaxies) is a scale-independent contribution arising from the fact that we use discrete tracers of the matter density field, such as galaxies. The template for the BAO analysis is generated with Eq.~\eqref{eq:Pkmu}.

As shown in Eq.~\eqref{eq:Pkmu}, it is clearer to express the anisotropic power spectrum as function of $k$ and $\mu$, instead of $k_\perp$ and $k_\parallel$. The rescaling  of distances appearing in Eq.~\eqref{eq:scaling} can be transformed to $k$ and $\mu$ as~\cite{Ballinger_scaling96}
\begin{equation}
\begin{split}
& k^{\rm true} = \frac{k^{\rm obs}}{\alpha_\perp}\left[1+\left(\mu^{\rm obs}\right)^2\left(F_{\rm AP}^{-2}-1\right)	\right]^{1/2}, \\
& \mu^{\rm true} = \frac{\mu^{\rm obs}}{F_{\rm AP}}\left[1+\left(\mu^{\rm obs}\right)^2\left(F_{\rm AP}^{-2}-1\right)	\right]^{-1/2},
\end{split}
\label{eq:scaling_kmu}
\end{equation}
where $F_{\rm AP} \!\equiv\! \alpha_\parallel/\alpha_\perp$.

Given the large scales probed, the line of sight changes with each pointing and cannot be considered parallel to any Cartesian axis.  Hence, it is not possible to obtain a well-defined $\mu$ for the observations, which makes a direct measurement $P(k,\mu)$ impossible. However, one can directly measure the Legendre multipoles of the anisotropic power spectrum using, e.g., the Yamamoto estimator~\cite{YamamotoEstimator}.\footnote{There are other compression options, such as the so-called angular wedges~\cite{Kazin_wedges}.} Then, the \textit{observed} power spectrum multipoles are modeled as
\begin{equation}
\begin{split}
& P_{\ell}(k^{\rm obs}) =  \frac{2\ell+1}{2\alpha_\perp^2\alpha_\parallel}\times \\
& \times \int_{-1}^{1}{\rm d}\mu^{\rm obs} P(k^{\rm true},\mu^{\rm true})\mathcal{L}_{\ell}(\mu^{\rm obs}) + A_\ell(k),
\end{split}
\label{eq:multipole_scale}
\end{equation}
where $\mathcal{L}_{\ell}$ is the Legendre polynomial of degree $\ell$, $P(k^{\rm true},\mu^{\rm true})$ is computed using Eqs.~\eqref{eq:Pkmu} and \eqref{eq:scaling_kmu}, and a $\left( r_{\rm d}^{\rm fid}/r_{\rm d}\right)^3$ term  has been absorbed into the constant factor $B$ of $P_{\rm sm}$ in Eq.~\eqref{eq:Psmk}. Different polynomials $A_\ell(k)$ are added to each one of the power spectrum multipoles.  These polynomials are added not only to marginalize over uncertainties related with non-linear clustering, but in particular to account for the possibility that the broadband of the template $P_{\rm m,sm}$  does not match the actual one.  These polynomials  have the form
\begin{equation}
A_\ell(k) = a_{\ell,1}k^{-3} + a_{\ell,2}k^{-2} + a_{\ell,3}k^{-1} + a_{\ell,4} + a_{\ell,5}k^{n},
\label{eq:Aell}
\end{equation}
where $n=1$ and $n=2$ before and after density field reconstruction, respectively~\cite{Beutler_BAODR12}. 
 
 Note that the standard BAO analysis can be performed with slightly different models of the signal that incorporate, for example, different nonlinear evolution and procedures to isolate the BAO feature.  These slight differences are discussed in Ref.~\cite{Hinton_BARRY} and do not affect our results or conclusions.

In summary, BAO-only analyses include the following parameters:
\begin{equation}
\left\lbrace \alpha_\perp, \alpha_\parallel, B, \beta, \bm{a}_\ell,\sigma_{\rm FoG},\Sigma_\perp,\Sigma_\parallel \right\rbrace,
\label{eq:params}
\end{equation}
where $\bm{a}_\ell$ are the coefficients of $A_\ell$ in Eq.~\eqref{eq:Aell}. All but the two first parameters $\alpha_\perp$ and $\alpha_\parallel$ are nuisance parameters. There are thus 17 parameters in the analysis of the monopole and quadrupole of the power spectrum (22 parameters if the hexadecapole is also included). In this work we consider the monopole and quadrupole of the post-reconstruction galaxy power spectrum. The values of the parameters listed in Eq.~\eqref{eq:params} that we use to compute the template are: $\alpha_\perp=\alpha_\parallel=1$, $B=b_{\rm g}^2$, $\beta=f/b_{\rm g}$, $\bm{a}_\ell = 0$, $\sigma_{\rm FoG} = 10$~Mpc$/h$, $\Sigma_\perp = 2$ Mpc$/h$ and $\Sigma_\parallel = 4$ Mpc$/h$. The value of $b_{\rm g}$ depends on the survey considered, $f$ depends on the redshift and cosmology assumed,  and we choose the values for $\Sigma_\perp$ and $\Sigma_\parallel$ following Ref.~\cite{Eisenstein_BAOrobust}. The specific values chosen for the fiducial parameter do not change our results.

\subsection{Covariance and likelihood}
Galaxy surveys normally rely on an estimation of the variation of the power spectrum using a suite of mock catalogs (see e.g.,~\cite{Kitaura_Patchy,White_qpm}) computed on the fiducial cosmology to obtain the covariance of the multipoles of the power spectrum. Using galaxy mocks makes it easier to model  selection effects, observational mask, survey geometry, and other observational systematics that need to be taken into account to avoid a bias in the measurement~\cite{Ross_BOSSsyst}. However, motivated by the ever growing number of mocks required to lower the noise of the covariance below the statistical errors, there are proposals to obtain precise covariance matrices directly from the data  (see e.g.,~\cite{Oconnell_covmat_nomocks}). Since in this work we aim to treat BAO analyses in general and do not need exquisite precision on the covariance, we use an analytic approximation. Neglecting mode coupling due to the non-linear collapse and the observational mask, we approximate the covariance per $k$ and $\mu$ bin as
\begin{equation}
 \sigma^2(k,\mu) = \frac{P^2(k,\mu)}{N_{\rm modes}(k,\mu)},
\label{eq:covmat_kmu}
\end{equation}
 where $N_{\rm modes}$ is the number of modes per bin in $k$ and $\mu$ in the observed volume $V_{\rm obs}$, given by
\begin{equation}
N_{\rm modes}(k,\mu) = \frac{k^2\Delta k\Delta\mu}{8\pi^2}V_{\rm obs},
\label{eq:Nmodes}
\end{equation}
where $\Delta k$ and $\Delta\mu$ are the widths of the $k$ and $\mu$ bins, respectively. Note that $P_{\rm shot}$ is implicitly included in Eq.~\eqref{eq:covmat_kmu}, since it is part of the modeling of $P(k,\mu)$ (see Eq.~\eqref{eq:Pkmu}). In the case of shot-noise subtracted power spectrum measurements, $P_{\rm shot}$ should be removed from Eq.~\eqref{eq:Pkmu} and added explicitly to $P(k,\mu)$ in Eq.~\eqref{eq:covmat_kmu}.

Once we have defined the covariance per $k$ and $\mu$ bin, we can compute the covariance of the power spectrum multipoles. This covariance can be decomposed in  sub-covariance matrices for each multipole, plus the sub-covariance matrices between different multipoles. For multipoles $\ell$ and $\ell^\prime$, the sub-covariance matrix under the Gaussian assumption is given by (see Ref.~\cite{Grieb_covmat} for a detailed derivation)
\begin{equation}
\begin{split}
\mathcal{C}_{\ell\ell^\prime}(k_i, k_j) & = \delta_{ij} \frac{(2\ell +1)(2\ell^\prime+1)}{2}\times
\\
& \times\int_{-1}^1{\rm d}\mu \sigma(k_i,\mu)^2\mathcal{L}_\ell(\mu)\mathcal{L}_{\ell^\prime}(\mu)\,,
\end{split}
\label{eq:covariance}
\end{equation}
where $\delta_{ij}$ is the Kronecker delta. Finally, it is necessary to incorporate observational effects (in both the clustering summary statistic and its covariance), which are mainly due to the geometry of the survey itself, in the theoretical modeling of the observed multipoles of the power spectrum. The geometry of the survey (the footprint and the fact that some regions may be observed more times than others) affects the selection of galaxies and is modeled using a mask in configuration space (which becomes a convolution in Fourier space). While modeling and properly implementing the mask (see Refs.~\cite{Wilson_mask,Beutler_BOSSDR12_BAO,Beutler_BOSSDR12_RSD} for detailed discussions) are key for obtaining reliable conclusions from power spectrum measurements, they do not  have an effect on the focus of our study: the flexibility of the analysis and its validity regarding cosmologies beyond $\Lambda$CDM at the level of perturbations. Moreover, the effect of the mask on the BAO analysis is small and more important for redshift-space-distortion measurements. Therefore, we do not account for observational effects in this work.

Taking all this into account, we can compare the theoretical prediction of the measured power spectrum with respect to a fiducial cosmology with actual observations. Hence, the corresponding likelihood is
\begin{equation}
\log L\propto \left(\bm{P}^{\rm th}-\bm{P}^{\rm data}\right)\mathcal{C}^{-1}\left(\bm{P}^{\rm th}-\bm{P}^{\rm data}\right)^T,
\label{eq:likelihood}
\end{equation}
where $\bm{P}$ denotes the concatenation of all the multipoles of the power spectrum considered in the analysis, $\bm{P}^{\rm th}$ and $\bm{P}^{\rm data}$ are the theoretical prediction and the measured power spectrum, respectively, the superscript $T$ refers to the transpose operator, and all quantities are evaluated at the observed values of $k$ (following Eq.~\eqref{eq:multipole_scale}).

\section{Estimating the bias on BAO measurements}
\label{sec:bias_estimation}
In this section we explore the impact that extensions to $\Lambda$CDM have on the BAO feature. After introducing the models considered, we study the changes in $O_{\rm lin}$ with respect to the fiducial cosmology, individually varying different parameters. Afterwards, we perform mock BAO analyses as detailed in Section~\ref{sec:analysis} to evaluate the flexibility and validity of BAO results for the extended cosmologies.

\subsection{Cosmological models}\label{sec:models}
As a point of reference, we consider $\Lambda$CDM with one massive neutrino species ($m_\nu=0.06$ eV) and its standard parameters: the physical baryon and cold dark matter density parameters today, $\Omega_{\rm b}h^2$ and $\Omega_{\rm cdm}h^2$, respectively, the reduced Hubble constant $h$, the spectral index $n_{\rm s}$, and the amplitude of the primordial power spectrum of scalar modes $A_{\rm s}$.
We take the fiducial cosmology in this section to be the same used for our analysis in Section~\ref{sec:bias_alphas}: $\Lambda$CDM with parameter values listed in Table~\ref{tab:models}. In addition to $\Lambda$CDM, we focus on two models that have been suggested as potential solutions of the Hubble tension: EDE and DNI.  We also consider a $\Lambda$CDM model with a different number of relativistic species than the prediction of the standard model $N_{\rm eff}=3.046$, $\Lambda$CDM$+N_{\rm eff}$, where  $N_{\rm eff}$ is an additional parameter of this model. 
 
The EDE model proposes the existence of a scalar field $\phi$ that is initially frozen (due to Hubble friction) at some field value and becomes dynamical after the Hubble parameter drops below some critical value. If the EDE becomes dynamical before recombination, it leads to a decrease in size of the photon-baryon sound horizon, providing a resolution to the Hubble tension \cite{Poulin_EDE,Smith_EDE}. Therefore, an EDE resolution to the Hubble tension leads to a decrease in $r_{\rm d}$, while keeping CMB power spectra angular scales and peak heights fixed via small shifts in the standard $\Lambda$CDM parameters. 
In particular we explore the oscillating EDE model, described in Ref.~\cite{Smith_EDE}. The scalar field evolves along an axion-like potential $V$ of the form 
\begin{equation}
V(\phi) = \Upsilon^4 \left[1-\cos\left(\phi/\lambda\right)\right]^{n_{\rm axion}},
\end{equation}
where $\Upsilon$ is the normalization of the potential and $\lambda$ is the decay constant of the scalar field. 
The additional parameters beyond those of $\Lambda$CDM considered for this model are the critical redshift $z_{\rm EDE}^c$ when the EDE becomes dynamical, the fraction $f_{\rm EDE}$ of EDE in the total energy density of the Universe at the critical redshift, the initial scalar field displacement $\Theta_{\rm EDE}\equiv \phi_i/\lambda$, and the power $n_{\rm axion}$ of the oscillating factor of the  potential for the scalar field.  The shift in $r_{\rm d}$ is mainly controlled by $f_{\rm EDE}$, with a weaker dependence on $z_{\rm EDE}^c$. The shape of the potential (which is controlled by $n_{\rm axion}$) determines the rate at which the EDE dilutes after $z_{\rm EDE}^c$, and $\Theta_{\rm EDE}$ mainly controls the evolution of the EDE perturbations (see also~\cite{Lin:2019qug, Smith_EDE}). 

The DNI model proposes an interaction between neutrinos and a small fraction $f_{\rm DNI}$ of the dark matter $\chi$ with mass $m_\chi$ in order to solve the $H_0$ tension \cite{Ghosh_DNIH0}. The scattering cross section $\sigma_{\chi\nu}$, taken to be independent of the neutrino temperature, is parametrized as $u_{\rm DNI} \propto \sigma_{\chi\nu}/m_\chi$. In this model, rather than decreasing the size of the sound horizon, the new interaction inhibits the free-streaming of neutrinos, which lowers the standard phase shift $\varphi$ induced by neutrinos in the BAO. The peak structure of the CMB power spectra can be roughly described by the position of the $p$-th peak
\begin{equation}
\ell_{p, {\rm peak}} \simeq (p \pi- \varphi) \frac{D_M(z_*)}{r_*},
\end{equation}
where the subscript $*$ denotes quantities at the time of recombination. 
In this model, $r_*$ does not change with respect to $\Lambda$CDM,\footnote{Although the redshift of last scattering $z_*$ and the redshift of the end of radiation drag $z_d$ do not coincide, we expect minimal model dependence in the difference between $r_*$ and $r_{\rm d}$. Therefore, we consider them as encapsulating redundant information.}  so that $D_M(z_*)$ needs to compensate the reduction of the phase shift in order to reproduce the measured location of the CMB peaks. Given that the physical matter, radiation, and cosmological constant energy densities are very constrained by other CMB features, the required decrease in $D_M(z_*)$ is achieved by increasing $H_0$. Since this model modifies the $\Lambda$CDM neutrino-induced phase shift in the BAO and the standard BAO analysis does not account for such effects, DNI might lead to a significantly biased BAO constraint. 
 Note that, as indicated in Ref.~\cite{Ghosh_DNIH0}, this model assumes massless neutrinos (with no implementation for massive neutrinos). Therefore, we will also consider massless neutrinos when assuming DNI cosmologies.

 We use the public Boltzmann code CLASS~\cite{Lesgourgues:2011re,Blas:2011rf},\footnote{\url{http://class-code.net/}} as well as its modifications to include the EDE and DNI\footnote{\url{https://github.com/subhajitghosh-phy/CLASS_DNI}} models, to compute the matter power spectra and other cosmological quantities needed. 
 
\subsection{Cosmological dependence of the shape of $\pmb{O_{\rm lin}}$ }

\begin{figure*}
\centering
\includegraphics[width=0.325\textwidth]{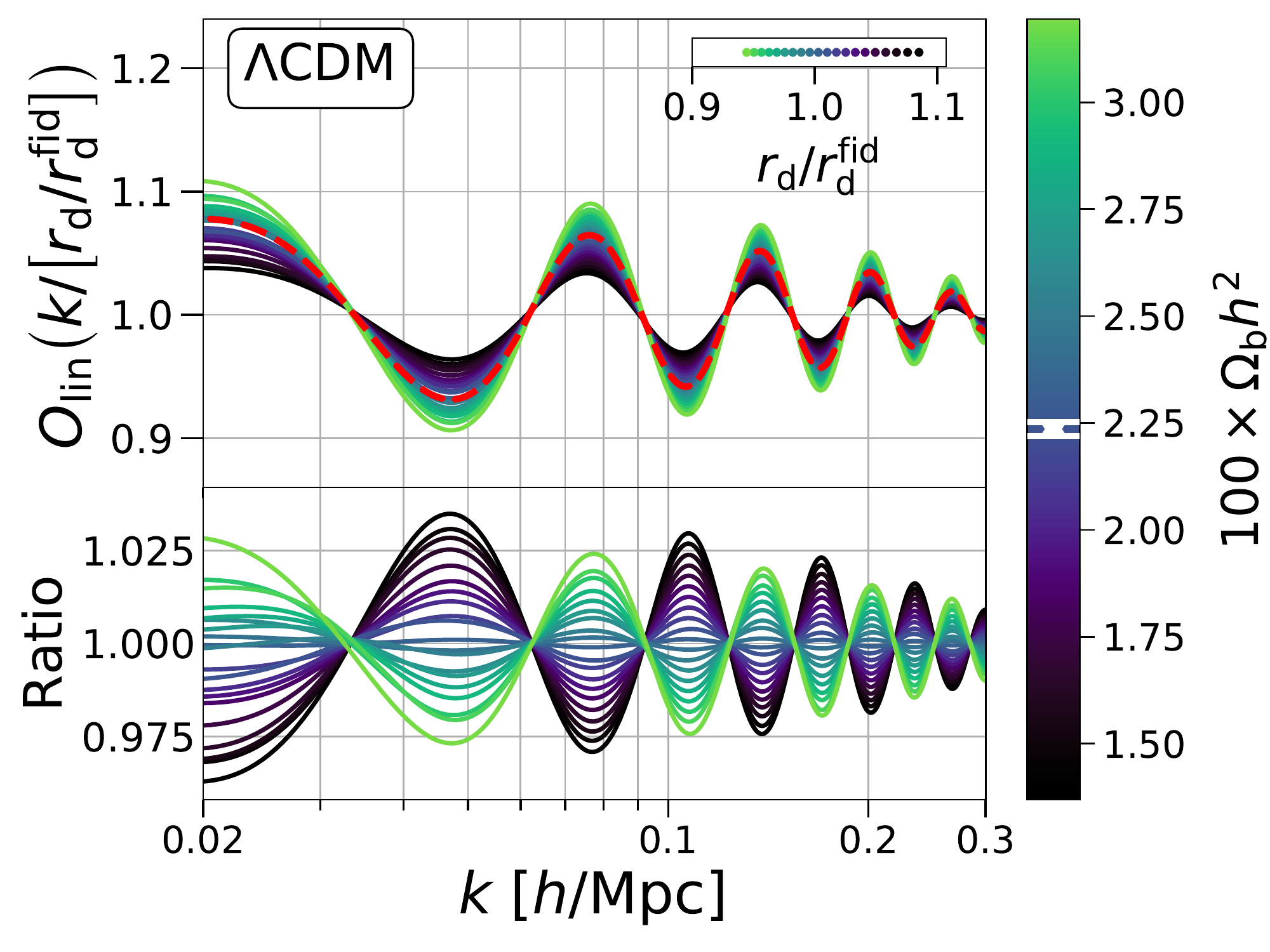}
\includegraphics[width=0.325\textwidth]{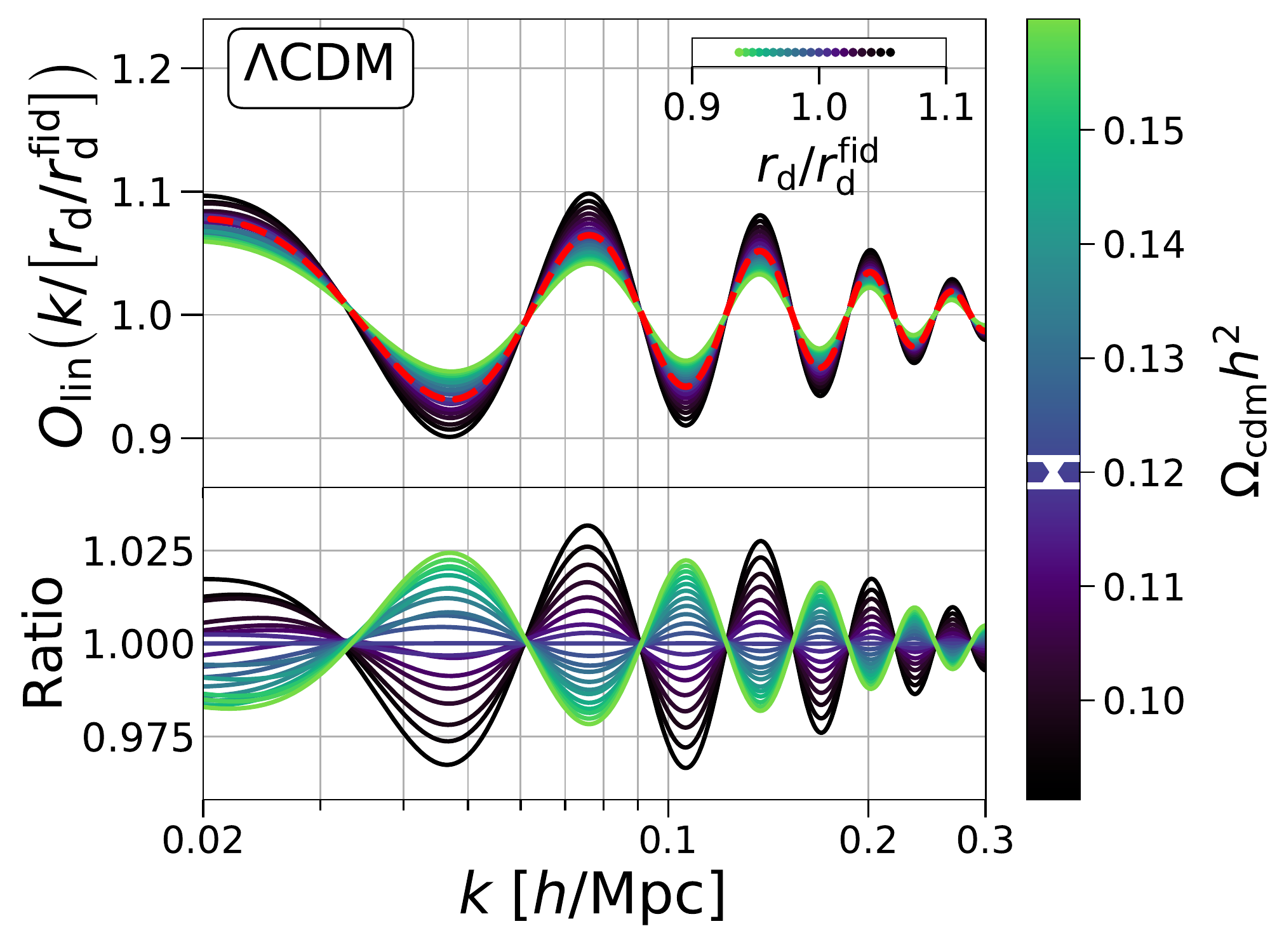}
\includegraphics[width=0.325\textwidth]{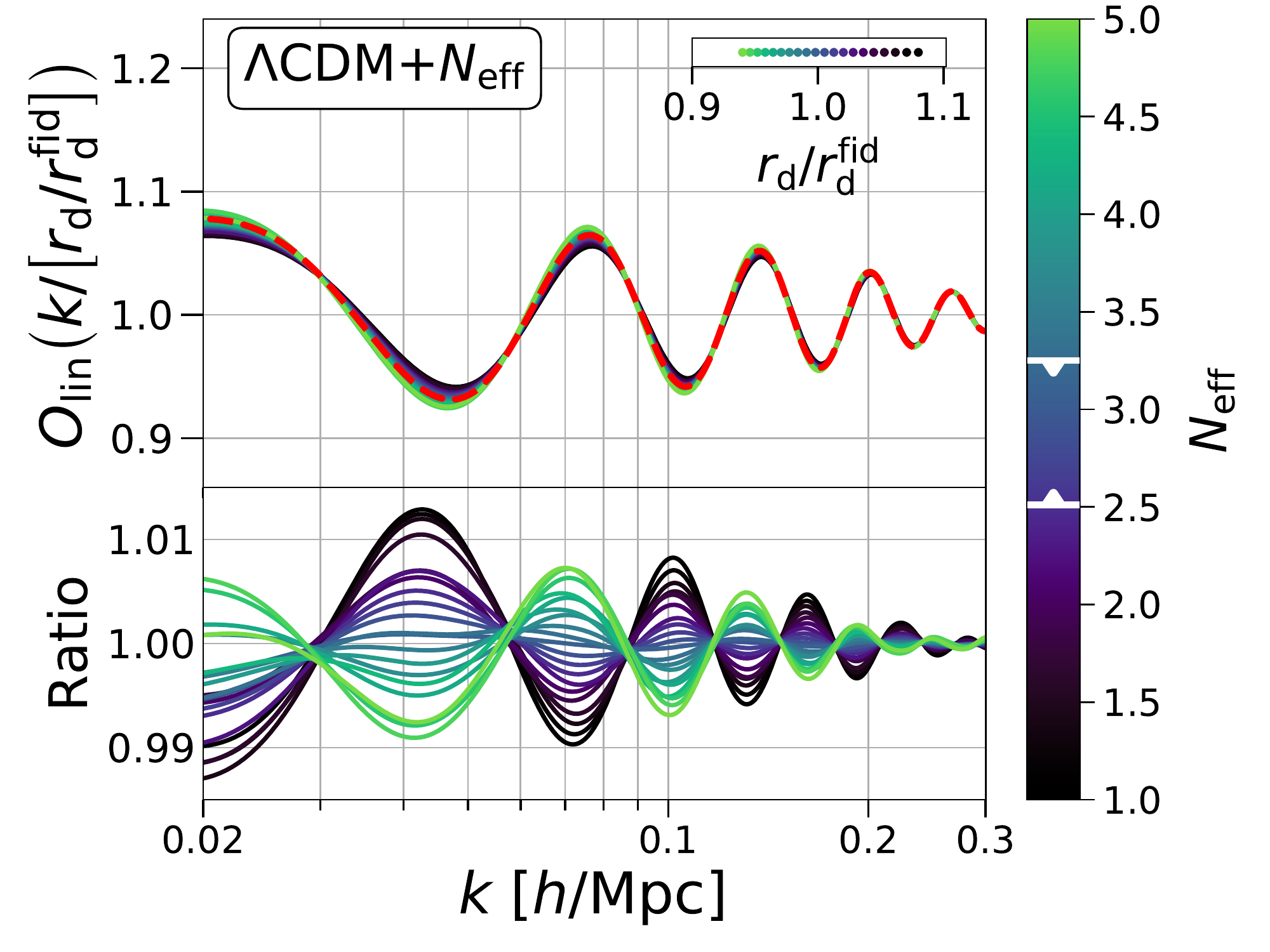}
\includegraphics[width=0.325\textwidth]{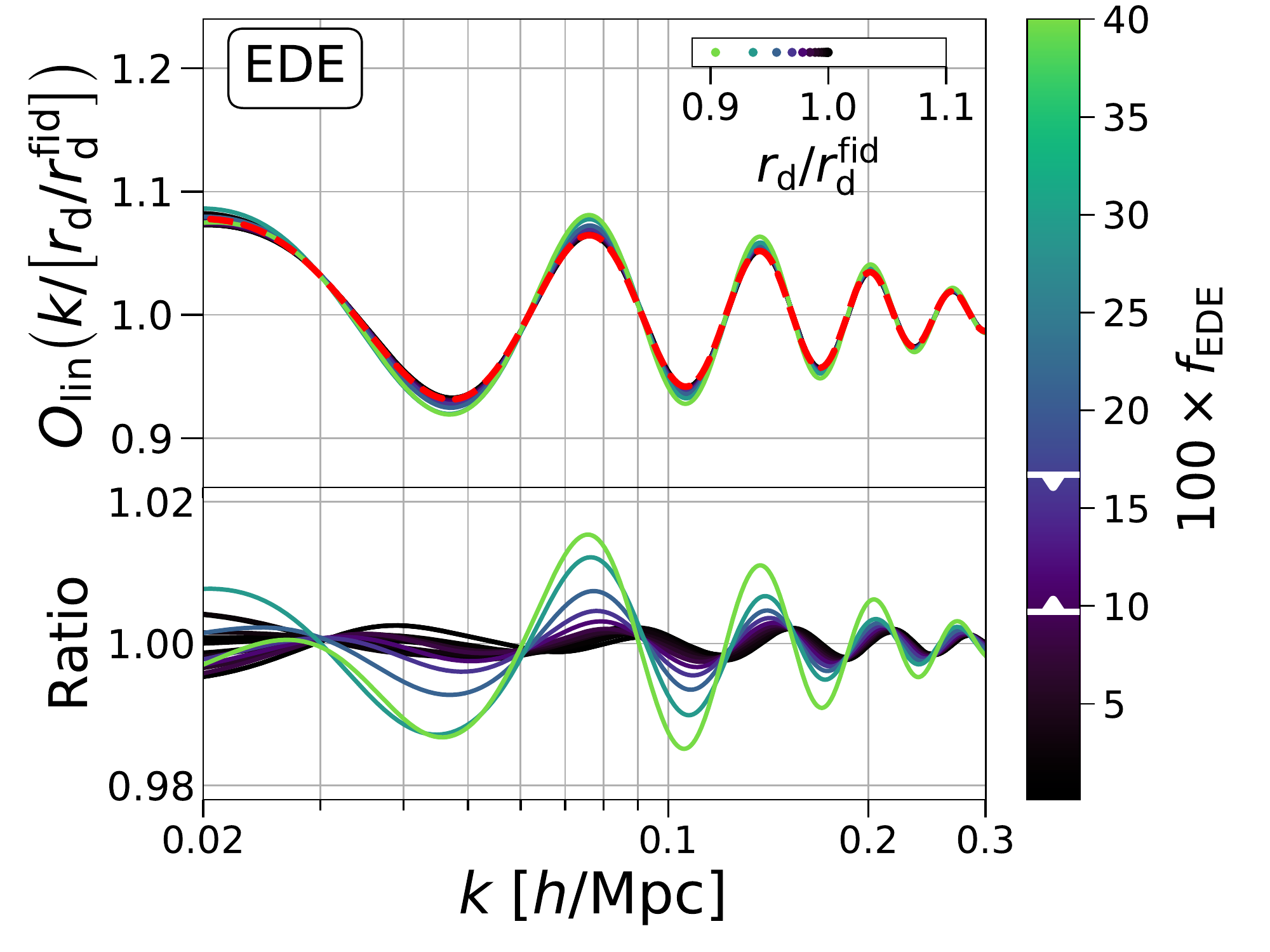}
\includegraphics[width=0.325\textwidth]{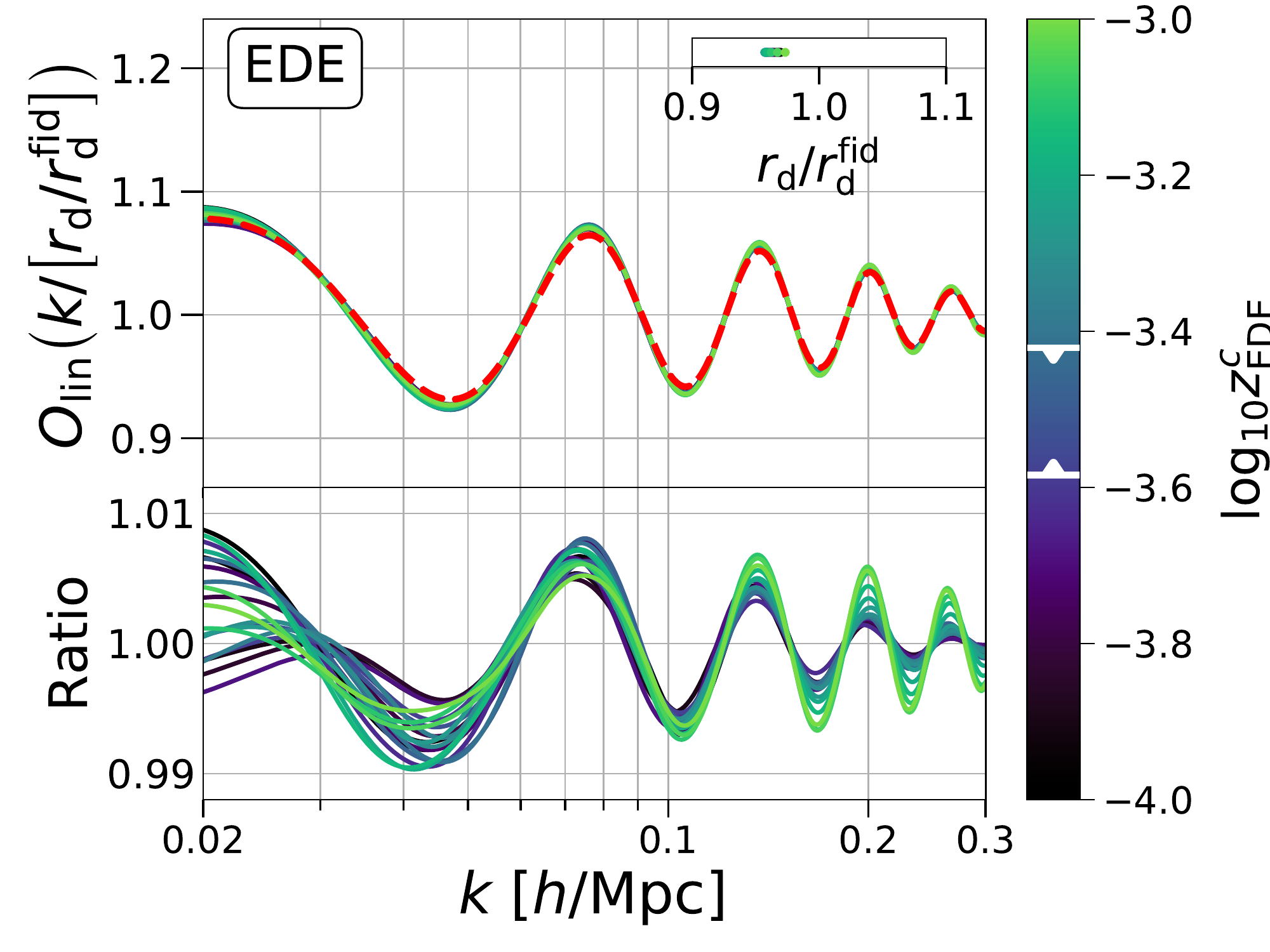}
\includegraphics[width=0.325\textwidth]{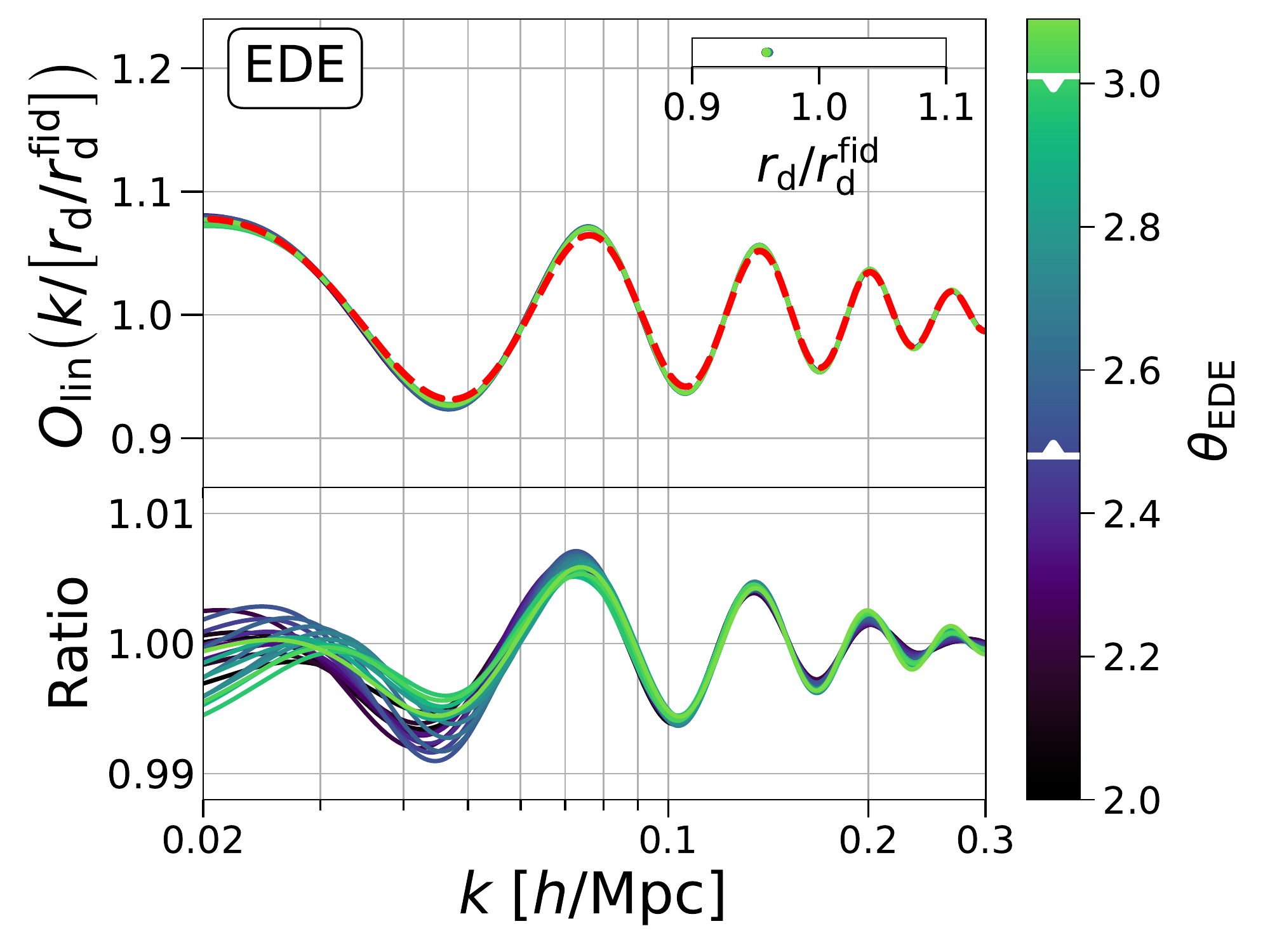}
\includegraphics[width=0.325\textwidth]{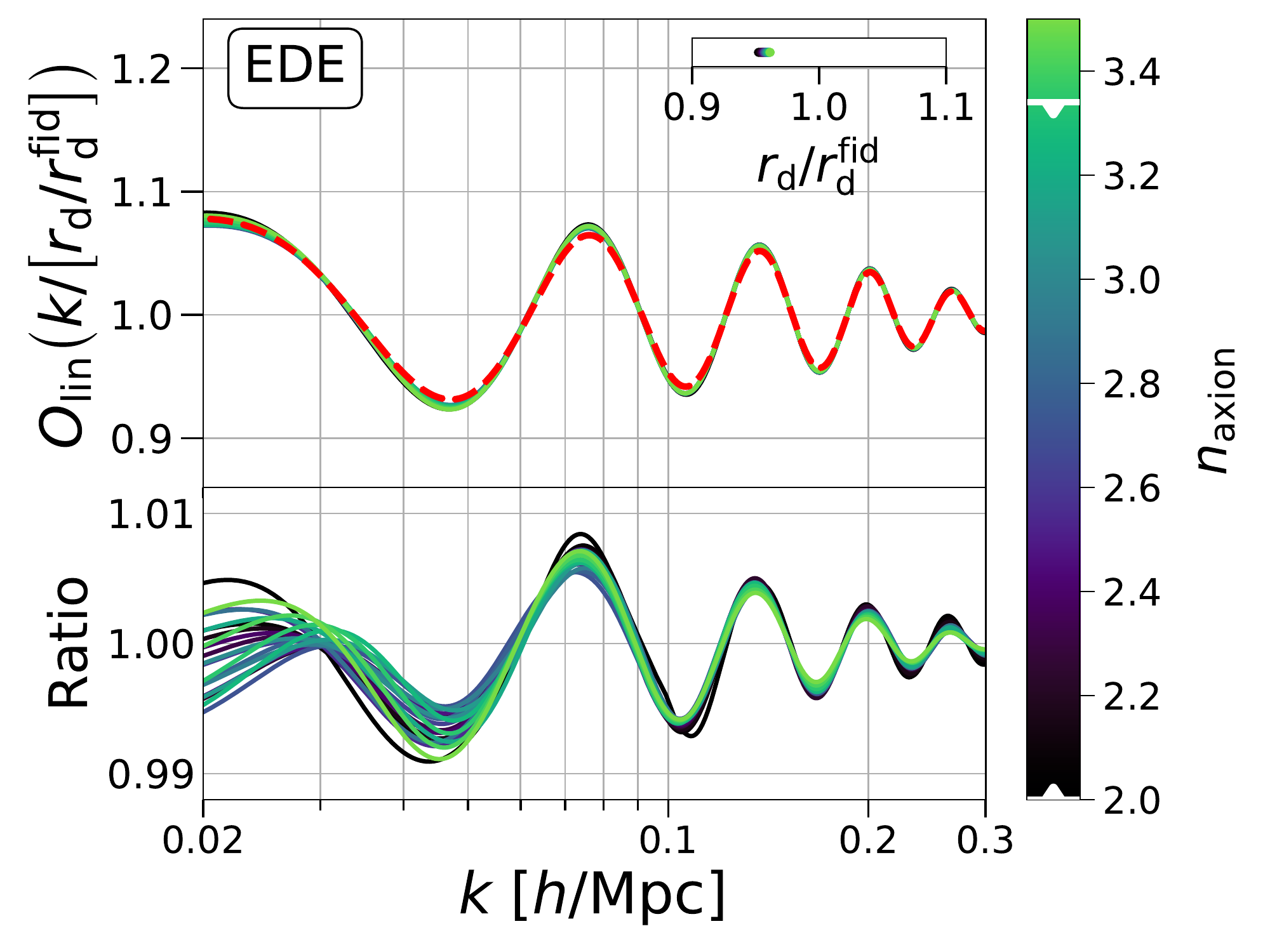}
\includegraphics[width=0.325\textwidth]{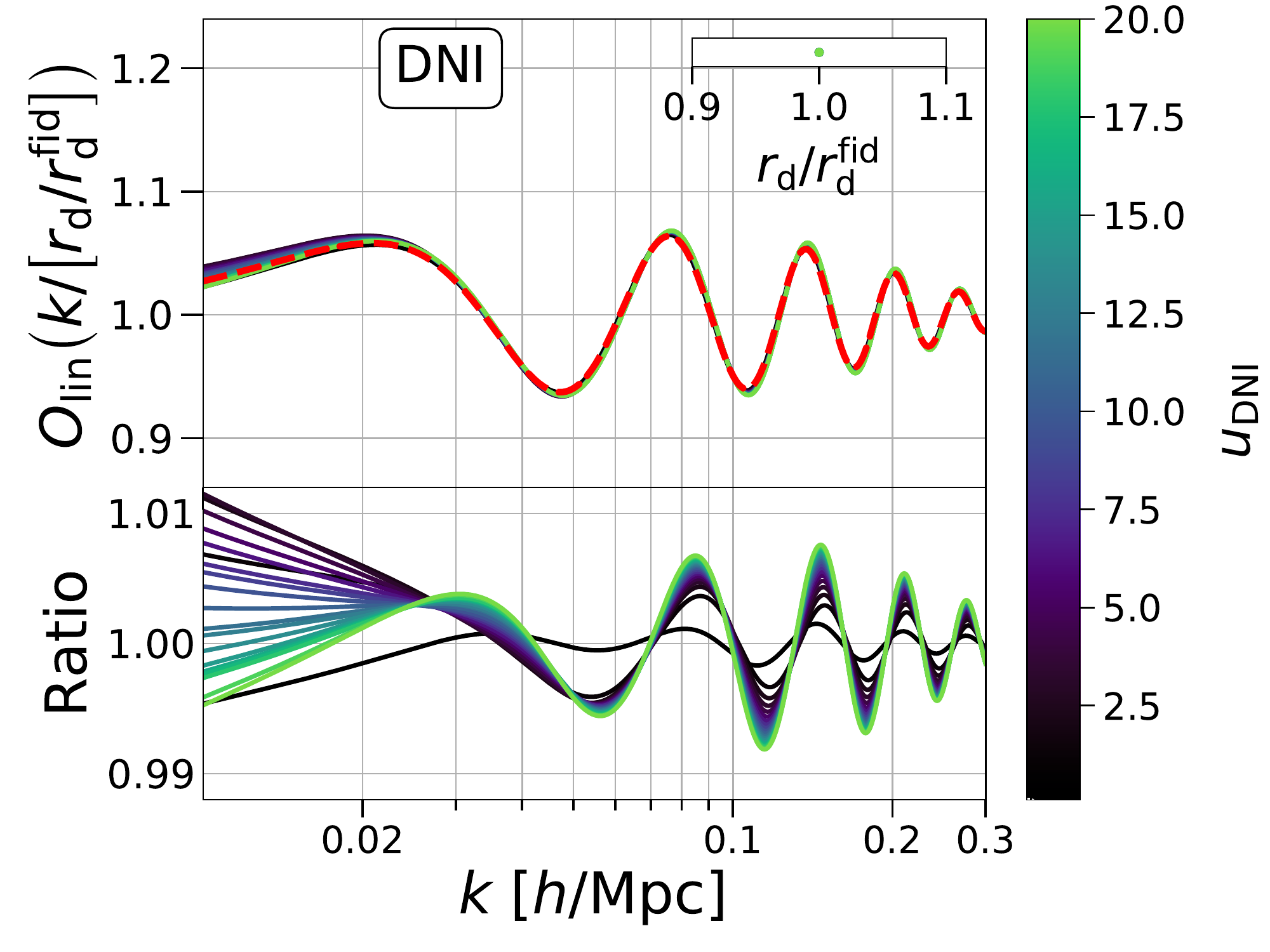}
\includegraphics[width=0.325\textwidth]{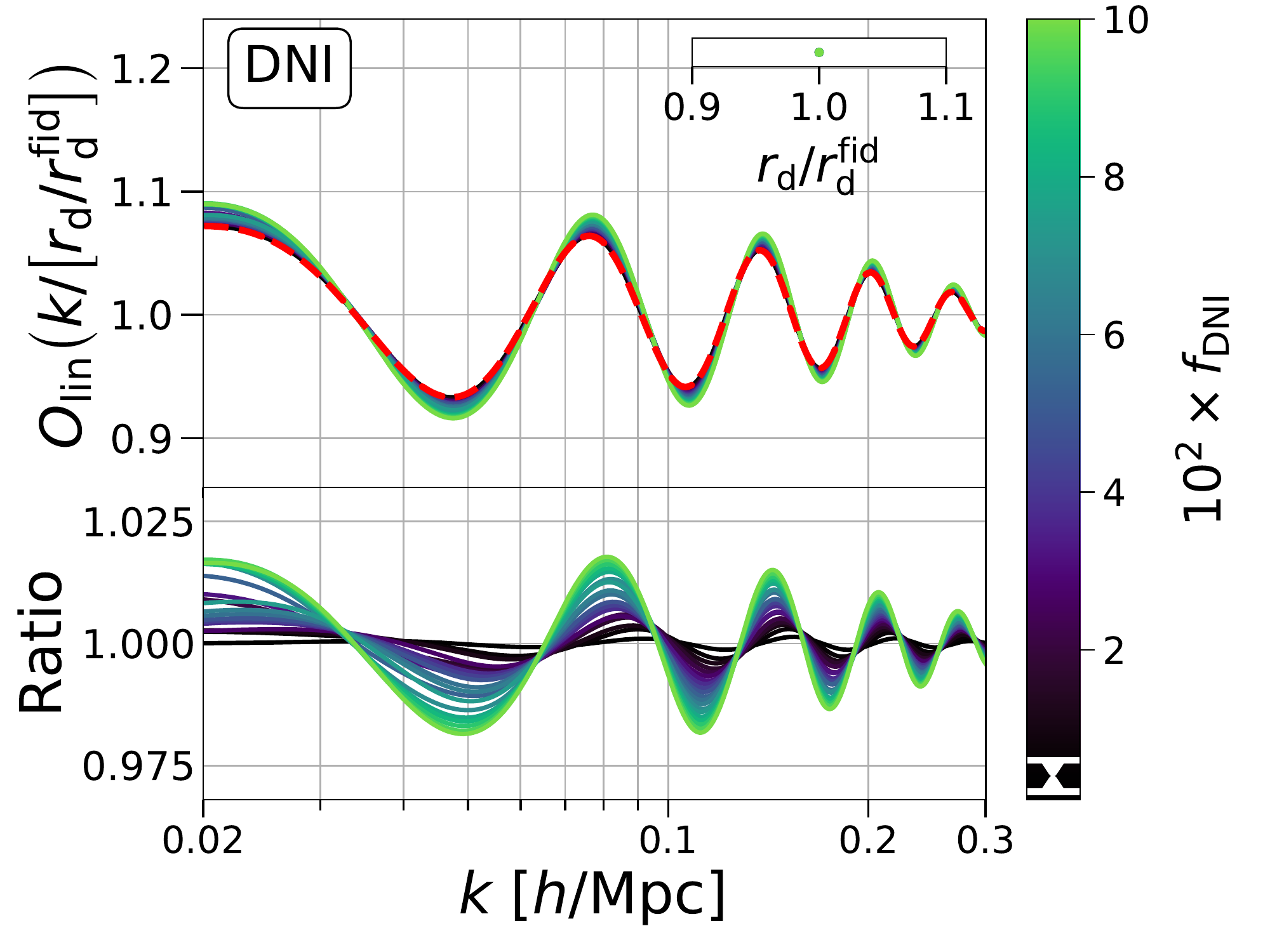}
\caption{BAO feature $O_{\rm lin}$ in the matter power spectrum for varying cosmological parameters (with $k$ rescaled by $r_{\rm d}/r_{\rm d}^{\rm fid}$), compared with the fiducial prediction (red dashed lines), and the corresponding ratio (i.e., modified over fiducial cosmology) in the lower panels. The insets show $r_{\rm d}/r_{\rm d}^{\rm fid}$ for each case, and the model under consideration is given in the upper left corner of each panel. We keep all $\Lambda$CDM parameters to their fiducial values except for when each of them is varied, as indicated in the corresponding color bars. The intervals limited by white lines in the color bars correspond to the 68\% confidence level constraints of each parameter from \textit{Planck} observations, and additional data sets for EDE and DNI. For the panels corresponding to EDE, we use $f_{\rm EDE}=0.2$, $\log_{10}z_{\rm EDE}^c=3.5$, $\Theta_{\rm EDE}=2.8$, and $n_{\rm axion}=3$, unless otherwise indicated. For the DNI model, we use $f_{\rm DNI}=0.02$ and $u_{\rm DNI}=5$, unless otherwise indicated. DNI constraints are reported in terms of $f_{\rm DNI}u_{\rm DNI}$, hence they are adapted to the fixed values of $u_{\rm DNI}$ and $f_{\rm DNI}$ assumed in this figure. Note the change in scale of the $y-$axis for the lower sections of each panel.}
\label{fig:Olin_rescaled}
\end{figure*}

We now consider how the BAO feature changes with respect to the fiducial under these different cosmological models with varying parameters.
In particular, we are concerned with changes in the evolution of perturbations before recombination, which affect $\alpha_\perp$ and $\alpha_\parallel$ through $r_{\rm d}$. Once we account for modifications of $r_{\rm d}$, we can investigate changes in the BAO feature beyond its characteristic scale. Changes to the background expansion history at low redshift affect $q_\perp$ and $q_\parallel$;  however, it has been shown that the standard BAO analysis is robust to such changes~\cite{Aubourg15,VargasMagana_BAOtheosyst, Carter_BAOtest}. Therefore, we do not model the Alcock-Paczynski effect nor the isotropic dilation  in this section and focus on the intrinsic undistorted shape of $O_{\rm lin}$. 

In order to study the BAO feature, we compare $O_{\rm lin}(k)$ predictions for the modified and fiducial cosmologies. If only the peak position shifts away from the fiducial, while the overall pattern remains the same, the result is a rescaling of $k$ in the argument of $O_{\rm lin}$ due to the variation of $r_{\rm d}$ (see Eq.~\eqref{eq:scaling}). In this case, all $O_{\rm lin}(k/[r_{\rm d}/r_{\rm d}^{\rm fid}])$ would be the same upon adjusting $r_{\rm d}$, where we have incorporated the $r_{\rm d}$ dependence in the rescaling of $k$ explicitly. If, however, the pattern of $O_{\rm lin}$ changes, then the standard BAO analysis does not sufficiently capture differences between the fiducial and the modified cosmologies. 

We show this comparison in Fig.~\ref{fig:Olin_rescaled}, where we individually vary all the relevant parameters of the cosmologies discussed in Section~\ref{sec:models}, keeping the rest of the parameters fixed. In order to study the variations thoroughly, we show both the absolute values of $O_{\rm lin}(k/[r_{\rm d}/r_{\rm d}^{\rm fid}])$ in the upper panels, as well as the ratio $O_{\rm lin}(k/[r_{\rm d}/r_{\rm d}^{\rm fid}])/\left(O_{\rm lin}(k/[r_{\rm d}/r_{\rm d}^{\rm fid}])\right)^{\rm fid}$ to compare with the fiducial in the lower panels. Note that at low redshift, the results shown in Fig.~\ref{fig:Olin_rescaled} are independent of redshift, since all redshift dependence is contained in $P_{\rm m,sm}$, except for the redshift evolution of the non-linear damping of the BAO, which we do not model in this work.

The first two panels of Fig.~\ref{fig:Olin_rescaled} show $O_{\rm lin}$, varying $\Omega_{\rm b}h^2$ and $\Omega_{\rm cdm}h^2$ under $\Lambda$CDM. We can see that $r_{\rm d}$ changes considerably for the parameter ranges explored (since the sound speed of the plasma and the matter content of the Universe change). However, after  rescaling  $r_{\rm d}$, the only significant change in $O_{\rm lin}$ is the amplitude of the wiggles, with no appreciable phase shift or further change of the BAO feature. Although the amplitude of the BAO is affected by non-linearities~\cite{Eisenstein_BAOrobust}, the non-linearities are modeled with an exponential decay dependent on $\Sigma_\perp$ and $\Sigma_\parallel$ (Eq.~\eqref{eq:Pkmu}), which we marginalize over. There might be some scale dependence of the BAO amplitude that is not covered by this modeling, but the  fact that there is no change in the position of the BAO implies that the constraints on $\alpha_\perp$ and $\alpha_\parallel$ should not be biased.

For $O_{\rm lin}$ under $\Lambda$CDM$+N_{\rm eff}$, shown in the third panel of Fig.~\ref{fig:Olin_rescaled}, the amplitude is also slightly modified, but there is also a phase shift introduced in the BAO wiggles. This phase shift can be seen in the skewness of the oscillations for a range of $N_{\rm eff}$~\cite{Baumann_NeffBAO_fisher}, as well as in the fact that the $k$-values where the ratio of $O_{\rm lin}$ (after rescaling $r_{\rm d}$) crosses 1 change with $N_{\rm eff}$. In addition, $N_{\rm eff}$ changes $r_{\rm d}$ significantly, due to the modification of the expansion history of the Universe before recombination. 

The next four panels of Fig.~\ref{fig:Olin_rescaled} show $O_{\rm lin}$ for EDE cosmologies, varying $f_{\rm EDE}$, $z_{\rm EDE}^{\rm c}$, $\Theta_{\rm EDE}$, and $n_{\rm axion}$.
The BAO feature is most sensitive to changes in $f_{\rm EDE}$. Although the amplitude of $O_{\rm lin}$ does not change significantly, there is a small phase shift introduced as $f_{\rm EDE}$ grows. A phase shift is also noticeable when varying the other EDE parameters, especially for $n_{\rm axion}$ (note that currently the whole range for $n_{\rm axion}$ covered by the colorbar is within the 68\% confidence level marginalized constraint reported in Ref.~\cite{Smith_EDE}). As noted in Ref.~\cite{Poulin_EDE}, varying $z_{\rm EDE}^c$ mainly affects the height of the first acoustic peak of the CMB power spectra  as well as the ratio $r_{\rm d}/r_{\rm Silk}$, where $r_{\rm Silk}$ is the Silk damping scale; thus, it has a noticeable effect on the amplitude ratio between different peaks. 

Finally, the last two panels of Fig.~\ref{fig:Olin_rescaled} show $O_{\rm lin}$ for DNI cosmologies, varying $u_{\rm DNI}$ and $f_{\rm DNI}$. The value of $r_{\rm d}$ remains the same, but the interaction between neutrinos and a fraction of the dark matter modifies the fiducial phase shift in the BAO feature. This change is larger for larger values of $u_{\rm DNI}$ or $f_{\rm DNI}$, evident from directly comparing curves of $O_{\rm lin}$ in the upper sections of the panels (and in e.g., Fig.~1 of Ref.~\cite{Ghosh_DNIH0} for the CMB power spectra). For this model, the ratio of $O_{\rm lin}$ with respect to the fiducial prediction may be misleading. Although the phase shift in $O_{\rm lin}(k/[r_{\rm d}/r_{\rm d}^{\rm fid}])$ is towards higher $k$ for larger values of $u_{\rm DNI}$ or $f_{\rm DNI}$, the amplitude of the BAO is also larger, causing the oscillations of the ratio of $O_{\rm lin}$ to shift towards lower $k$.

\renewcommand{\arraystretch}{1.3}
\begin{table*}[]
\centering
\resizebox{\textwidth}{!}{
\begin{tabular}{|c|c|c|c|c|c||c||c|c|c|c||c|c||c|}
\hline
 & $100\times \Omega_{\rm b}h^2$ & $\Omega_{\rm cdm}h^2$ & $h$ & $n_{\rm s}$ & $10^9\times A_{\rm s}$ & $N_{\rm eff}$ & $f_{\rm EDE}$ & $\log_{10}z_{\rm EDE}^c$ & $\Theta_{\rm EDE}$ & $n_{\rm axion}$ & $u_{\rm DNI}$ & $f_{\rm DNI}$ & $\Delta\chi^2$\\ \hline
Fiducial & 2.237 & 0.1200 & 0.6736 & 0.9649 & 2.100 & 3.046 & - & - & - & - & - & - & - \\ \hline
$\Lambda$CDM 1 & 2.229 & 0.1212 & 0.6680 & 0.9608 & 2.077 & 3.046 & - & - & - & - & - & -&  2.0 \\ \hline
$\Lambda$CDM 2 & 2.261 & 0.1160 & 0.6900 & 0.9709 & 2.140 & 3.046 & - & - & - & - & - & - & 18.8 \\ \hline
$\Lambda$CDM$+N_{\rm eff}$ 1 & 2.228 & 0.1154 & 0.6614 & 0.9587 & 2.072 & 2.796 & - & - & - & - & - & - &  1.3\\ \hline
$\Lambda$CDM$+N_{\rm eff}$ 2 & 2.287 & 0.1231 & 0.7061 & 0.9801 & 2.116 & 3.400 & - & - & - & - & - & - & 18.0 \\ \hline
EDE 1 & 2.251 & 0.1320 & 0.7281 & 0.9860 & 2.191 & 3.046 & 0.132 & 3.531 & 2.72 & 2.60 & - & - & 7.8\\ \hline
EDE 2 & 2.261 & 0.1413 & 0.7552 & 0.9928 & 2.221 & 3.046 & 0.200 & 3.545 & 2.43 & 2.34 & - & - & 50.1\\ \hline
DNI 1 & 2.253 & 0.1180 & 0.7023 & 0.9492 & 2.019 & 3.046 & - & - & - & - & 18.6 & $10^{-3}$&  6.4 \\ \hline
DNI 2 & 2.230 & 0.1212 & 0.7200 & 0.9600 & 2.111 & 3.046 & - & - & - & - & 15.0 & 0.02&  338\\ \hline
\end{tabular}
}
\caption{Cosmological models considered in this work and their corresponding cosmological parameter values. The `Fiducial' model is used to generate the template for the BAO fit. All other models are used as target cosmologies to generate the mock power spectra. Models denoted with `1' correspond to the good fits to \textit{Planck}, while models `2' are poor fits, as described in the text. Note that $\Omega_{\rm cdm}h^2$ represents the total dark matter density and includes the portion of dark matter that interacts with neutrinos in the DNI model. We find the minimum $\chi^2$ for each cosmology and the corresponding best-fit nuisance parameters using iMinuit.\footnote{\url{https://iminuit.readthedocs.io/en/stable/}}}
\label{tab:models}
\end{table*}
\renewcommand{\arraystretch}{1}

\subsection{Bias in $\pmb{\alpha_\perp}$ and $\pmb{\alpha_\parallel}$}\label{sec:bias_alphas}
Now that we have established that there are cosmological models that can produce $O_{\rm lin}$ patterns that do not match the fiducial prediction under a rescaling of $r_{\rm d}$ (or the BAO amplitude), we are ready to  evaluate the bias on the standard BAO fit that may arise under these cosmologies. In order to do so, we perform mock BAO analyses using the formalism described in Section~\ref{sec:analysis}, with the aim of evaluating the potential bias introduced in the determination of $\alpha_\perp$ and $\alpha_\parallel$. Such a bias could affect the constraints on the expansion history of the Universe and hence the inference of the cosmological parameters. Although it has been demonstrated~\cite{Aubourg15,VargasMagana_BAOtheosyst, Carter_BAOtest} that BAO-inferred distances are extremely robust to modifications on the expansion history of the Universe at late times,  these distances might not be robust against changes in the acoustic oscillations due to the growth of perturbations before recombination.

  Whether or not differences in the BAO pattern with respect to the fiducial prediction such as those highlighted in Fig.~\ref{fig:Olin_rescaled} introduce a potential bias on the BAO measurements depends on the flexibility of the BAO analysis. If  a bias were found for a given cosmological model, current BAO measurements (which assume a $\Lambda$CDM fiducial) would not be suitable for constraining that model. 
 
 In order to perform the mock BAO analysis, we create a mock power spectrum for each specific underlying cosmology under consideration, as if it were measured from the observed galaxy distribution.
Details on how we compute the mock power spectra can be found in Appendix~\ref{sec:mockdata}.
We perform the standard BAO analysis on each mock spectrum using the same fiducial $\Lambda$CDM cosmology. We run Monte Carlo Markov chains using the Monte Carlo sampler \texttt{emcee}~\cite{emcee}\footnote{\url{https://github.com/dfm/emcee}} to perform the BAO fit with the likelihood in Eq.~\eqref{eq:likelihood}.

Once we obtain the marginalized constraint in the $\alpha_\perp$-$\alpha_\parallel$ plane for each mock analysis, we can locate the point in parameter space at which the value of the marginalized posterior peaks to compare with the true point. We compute $\Delta\chi^2 = -2\Delta\log {L}$ between these two points  in the $\alpha_\perp$-$\alpha_\parallel$ plane assuming a Gaussian marginalized posterior and calculate the corresponding confidence level for a two-dimensional $\chi^2$ distribution.   This procedure is equivalent to computing the cumulative probability of the $\chi^2$ distribution up to the value of interest. Finally, we relate the calculated confidence level to a number of standard deviations $\sigma$ for a Gaussian distribution. 
We estimate the bias then to be this number of $\sigma$.

 In Table~\ref{tab:models}, we specify the values of the cosmological parameters for both the assumed fiducial cosmology needed for the BAO analysis and all of the target cosmologies used to generate the mock power spectra.  We use for the cosmological parameters of our  fiducial cosmology the mean values of the $\Lambda$CDM analysis (including temperature, polarization,  and lensing data) to \textit{Planck}~\cite{Planck18_pars}.  In addition, we consider two different sets of cosmological parameters for all models: the best fit to \textit{Planck} data (or \textit{Planck} data combined with other data sets)  and a poor fit. We refer to these two sets with the labels `1' and `2,' respectively, both individually (e.g., in the model name) and collectively.  For the latter, we choose one of the parameters to be at least $\gtrsim 3\sigma$ away from its nominal best-fit value and set the remaining parameters to their new best-fit values to \textit{Planck}, obtained from keeping the chosen parameter fixed.

  In addition to the the standard $\Lambda$CDM parameters and  $N_{\rm eff}$, we consider the extra parameters of EDE and DNI. For $\Lambda$CDM and $\Lambda$CDM$+N_{\rm eff}$, we use best-fit cosmologies obtained from \textit{Planck}'s publicly-available Monte-Carlo Markov chains,\footnote{\url{http://pla.esac.esa.int/pla/}} and obtain poor fits from the same chains fixing $H_0=69.00 \, {\rm km\,s^{-1}Mpc^{-1}} $ for $\Lambda$CDM and $N_{\rm eff}= 3.4$ for $\Lambda$CDM$+N_{\rm eff}$. The EDE 1 and DNI 1 cosmologies are taken from the reported best fits for the considered data sets in Refs.~\cite{Smith_EDE} and~\cite{Ghosh_DNIH0}, respectively. EDE 2 is obtained from the Monte-Carlo Markov  chain  computed for Ref.~\cite{Smith_EDE}, fixing $f_{\rm EDE}$ to 0.2, while DNI 2 is the best fit to \textit{Planck} data for fixed $u_{\rm DNI}=15$ and $f_{\rm DNI}= 0.02$, found using MontePython~\cite{Audren13_mp}.  The DNI 2 case has been chosen as a very extreme case to test the flexibility of the standard BAO analysis: its $\Delta\chi^2$ for \textit{Planck} power spectra with respect to the $\Lambda$CDM mean values (i.e., our fiducial case) is $\sim 338$. 

In order to mimic the power of current and next-generation galaxy surveys, we use  four different sets of survey specifications, inspired by BOSS~\cite{Alam_bossdr12} and DESI~\cite{desi}. These specifications are important, because they determine the effective redshift at which the power spectrum is measured and computed, as well as its covariance matrix. The specifications for these surveys are listed in Table~\ref{tab:surveys}.

\renewcommand{\arraystretch}{1.3}
\begin{table}[]
\centering
\resizebox{\columnwidth}{!}{
\begin{tabular}{|c|c|c|c|c|c|}
\hline
Survey & $z$ & $\Omega_{\rm sky}\, \left[{\rm deg}^2\right]$ & $V\, \left[\left({\rm Gpc}/h\right)^3\right]$ & $10^4 n_g\, \left[\left({\rm Mpc}/h\right)^{-3}\right]$ & $b_{\rm g}$ \\ \hline
BOSS LOWZ & 0.32 & 10000 & 1.27 & 2.85 & 1.85 \\ \hline
BOSS CMASS & 0.57 & 10000 & 3.70 & 2.10 & 1.85 \\ \hline
DESI 1 & 0.80 & 14000 & 6.80 & 12.3 & 1.86 \\ \hline
DESI 2 & 1.00 & 14000 & 8.73 & 6.41 & 1.50 \\ \hline
\end{tabular}
}
\caption{Specifications for each of the galaxy surveys considered in the mock BAO analyses. The columns give the effective redshift $z$ at which the power spectrum is measured, the sky coverage $\Omega_{\rm sky}$ of the survey, the physical volume $V$ covered by the survey, the mean galaxy density $n_g$, and the galaxy bias $b_{\rm g}$. }
\label{tab:surveys}
\end{table}
\renewcommand{\arraystretch}{1.}

\begin{figure*}[t]
\centering
\includegraphics[width=0.5\textwidth]{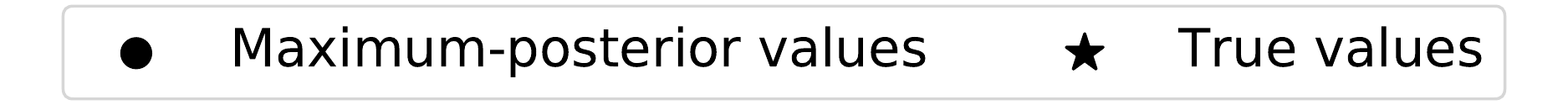}
\includegraphics[width=\textwidth]{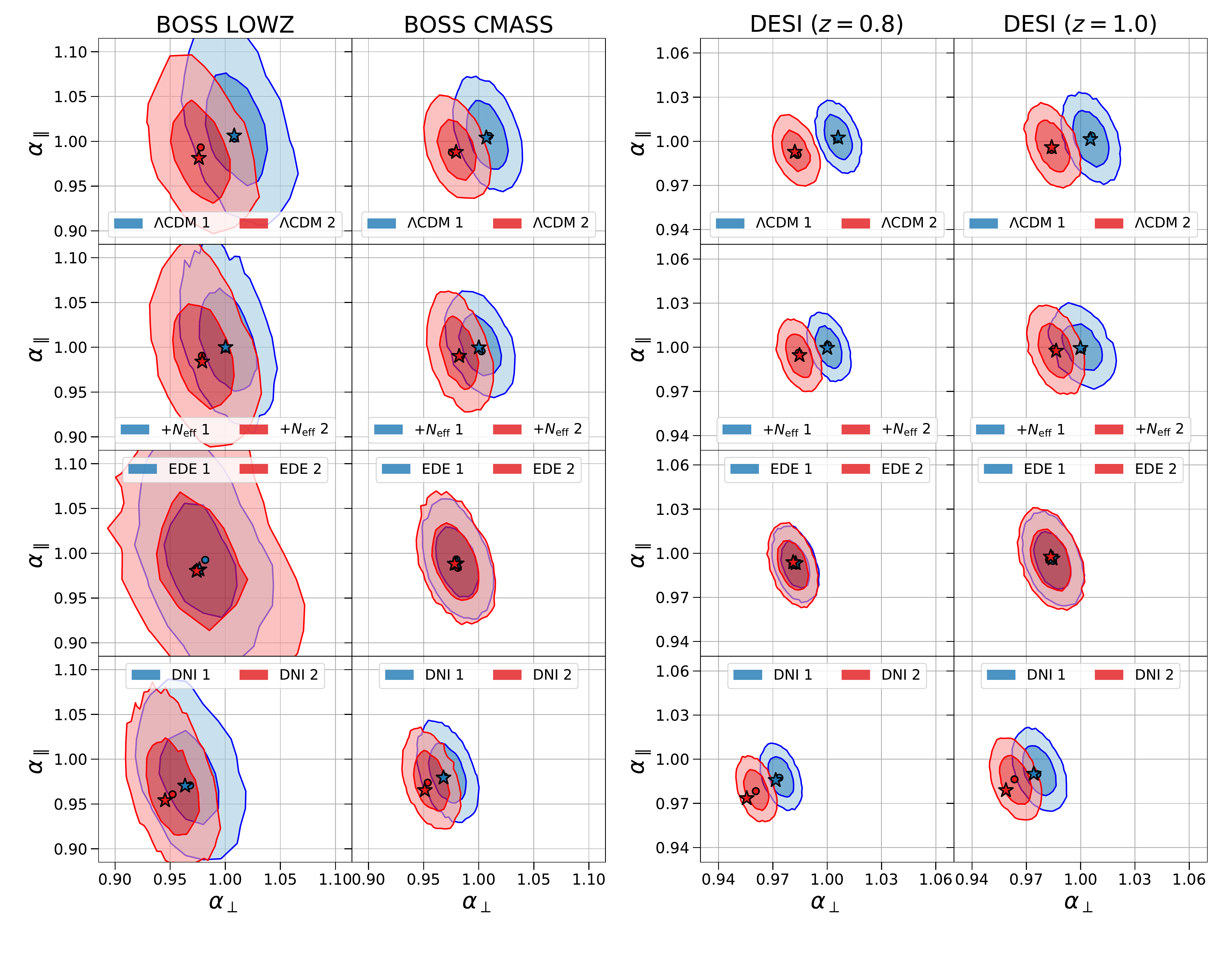}
\caption{The 68\% and 95\% confidence level marginalized constraints in the $\alpha_\perp$-$\alpha_\parallel$ plane, shown with their maximum-posterior and true values (represented as circles and stars, respectively), for different mock galaxy power spectra computed in $\Lambda$CDM, $\Lambda$CDM+$N_{\rm eff}$, EDE, and DNI, as indicated in the legend (rows), and for BOSS LOWZ and CMASS, as well as DESI at $z=0.8$ and $z=1.0$ (columns; note the different scales in the columns in the left and in the right). For all analyses, we use a template computed under a fiducial that matches the mean values of the $\Lambda$CDM parameters from \textit{Planck}'s analysis. Blue contours refer to the cosmologies numbered as `1' (best fit from \textit{Planck}, as well as external data sets for EDE and DNI) and red contours to cosmologies numbered as `2'. Cosmological parameters and survey specifications are listed in Tables~\ref{tab:models} and \ref{tab:surveys}, respectively.}
\label{fig:alphas}
\end{figure*}

Marginalized constraints in the $\alpha_\perp$-$\alpha_\parallel$ plane are shown in Fig.~\ref{fig:alphas}, comparing the values for which the posterior distribution peaks and the true values of these parameters (marked by  circles and  stars, respectively). Each panel shows the results for a given galaxy survey and cosmological model. We show  cosmologies 1 in blue and cosmologies 2 in red.  Note that, since each model has its own prediction for $D_M/r_{\rm d}$ and $Hr_{\rm d}$ (and these quantities change with redshift), each panel shows different true values for $\alpha_\perp$ and $\alpha_\parallel$.  We show results assuming finished surveys (BOSS~\cite{Alam_bossdr12}) in the two left columns, while the two right columns correspond to future surveys (DESI~\cite{desi}).

The true values of $\alpha_\perp$ and $\alpha_\parallel$ lie well within the 68\% confidence level contours in  Fig.~\ref{fig:alphas}, indicating that there is no significant bias in the inference of $\alpha_\perp$ and $\alpha_\parallel$ for any of the models in the finished surveys, even when the fiducial model and the model under study have very different predictions for $r_{\rm d}$ or $O_{\rm lin}$. Moreover, the largest bias found for our extreme DNI 2 example is $\sim 1\sigma$, for DESI 1 survey (and $\sim 0.8\sigma$ for DESI 2 survey). 
  This bias is introduced from the change in the phase shift produced by the interaction between neutrinos and dark matter, as demonstrated in Fig.~\ref{fig:Olin_rescaled}, and expected from the discussion in~\ref{sec:models}. However, note that for DNI 1, the bias in the BAO measurements is always $\lesssim 0.2\sigma$. 

Finally, in addition to biasing the best fit, inaccurate modeling can also induce a misestimation of the parameter uncertainties (see e.g,~\cite{Bellomo_Fisher}). However, we find that not only there is no significant bias in the best-fit values of $\alpha_\perp$ and $\alpha_\parallel$, but their uncertainties do not change. We can appreciate this when comparing the marginalized constraints shown in Fig.~\ref{fig:alphas} for cosmologies 1 and 2. In each panel, the uncertainties for each of the cases are the same. This means that the BAO standard analysis does not misestimate the covariance between $\alpha_\perp$ and $\alpha_\parallel$, even if the assumed fiducial cosmology is very far from the cosmology corresponding to the observations. 

\section{Discussion}\label{sec:discussion}
The results shown in Section~\ref{sec:bias_alphas} are clear and powerful. The BAO standard analysis is flexible enough to capture different growth-of-structure histories prior to recombination for the models explored in this work. Moreover, our results indicate that the inference of $\alpha_\perp$ and $\alpha_\parallel$ using a template  can be unbiased even for cosmologies that differ significantly from the fiducial cosmology used to compute the template. In most of the cases, this fiducial cosmology would come from the best fit to the CMB power spectra. 

These results, as well as the discussion included in Section~\ref{sec:info}, provide a strong motivation to keep using template-fitting methodologies to perform BAO-only analysis: this approach is robust, flexible, and general enough so that it allows agnostic studies of cosmology. Therefore, the results of standard BAO analyses can be used to infer cosmological parameters without introducing any bias in the subsequent constraints. 
Moreover, compressing the BAO information in terms of  $\alpha_\perp$ and $\alpha_\parallel$ reduces the probability of introducing observational systematics and reduces computational costs on cosmological parameter inference, which proves key as the amount of observations grows. 

In this work we have focused on the BAO fit to the galaxy power spectrum. However, other tracers can be used to infer the matter power spectrum; regardless of which tracer is used,  BAO imprints are present in the perturbations of all of them. Therefore, our study can be extrapolated to the power spectrum of the fluctuations of the number of quasars, the Lyman-$\alpha$ forest, line-intensity mapping, etc.,  as well as their cross-correlations. The particularities of the analysis applied to each specific tracer are related with observational systematics and not with the underlying cosmology; hence, they do not affect the extrapolation of the conclusions of this work. 

Finally, although we have focused on four specific cosmological models, our results can be extrapolated to other cosmological models. As explored in Appendix~\ref{sec:CMBtoBAO}, the BAO feature present on the matter density distribution at the time of recombination (indirectly probed by CMB observations to great precision) survives unmodified until the present. This is true modulo non-linear collapse of perturbations, which still preserves some of the features encoded in the BAO, as shown in Ref.~\cite{Baumann_NeffBAO}. 

Therefore, we conclude that using a template computed for a good fit to the CMB observations should be suitable for a BAO analysis. This assessment might not hold, however, if the evolution of dark matter and baryon overdensities is influenced by beyond-$\Lambda$CDM physics post recombination, such as baryon-dark matter interactions (e.g., see Refs.~\cite{Dvorkin:2013cea,Boddy:2018wzy} for a description of the modified Boltzmann equations). Moreover, general relativity is assumed in density-field reconstruction (see e.g.,~\cite{Eisenstein_reconstruction}), hence only BAO measurements obtained before applying density-field reconstruction shall be used to constrain cosmological models that modify gravity.

\subsection{Constraining cosmological models with BAO}
Cosmological parameter inference  using $\alpha_\perp$-$\alpha_\parallel$ constraints  would be completely accurate only if the BAO measurements are unbiased for every power spectrum computed at every point of the parameter space of the model. However, exploring these biases for the whole parameter space is unfeasible. We overcome this limitation by studying concrete cases far from the assumed fiducial cosmology.  These cases represent the tails of the posterior, where residual small biases in  the BAO constraints would not modify the bulk of the  distribution nor introduce a significant bias in the parameter inference.

 On the other hand, it is expected that the bias in $\alpha_\perp$ and $\alpha_\parallel$ decreases as we move in the  parameter space from the region corresponding to these cosmologies towards the  fiducial cosmology. Therefore, an unbiased analysis of the extreme cases that we consider further supports the use of $\alpha_\perp$ and $\alpha_\parallel$ constraints for joint parameter inference of cosmological models beyond $\Lambda$CDM with other cosmological probes, such as the CMB power spectra.  This means that our results indicate that previously reported BAO measurements in the form of $\alpha_\perp$ and $\alpha_\parallel$ constraints are valid for constraining not only late-time deviations of the expansion history of the Universe, but also early-Universe  extensions to $\Lambda$CDM, without the need to refit the galaxy power spectrum. 

\begin{figure}
\centering
\includegraphics[width=\columnwidth]{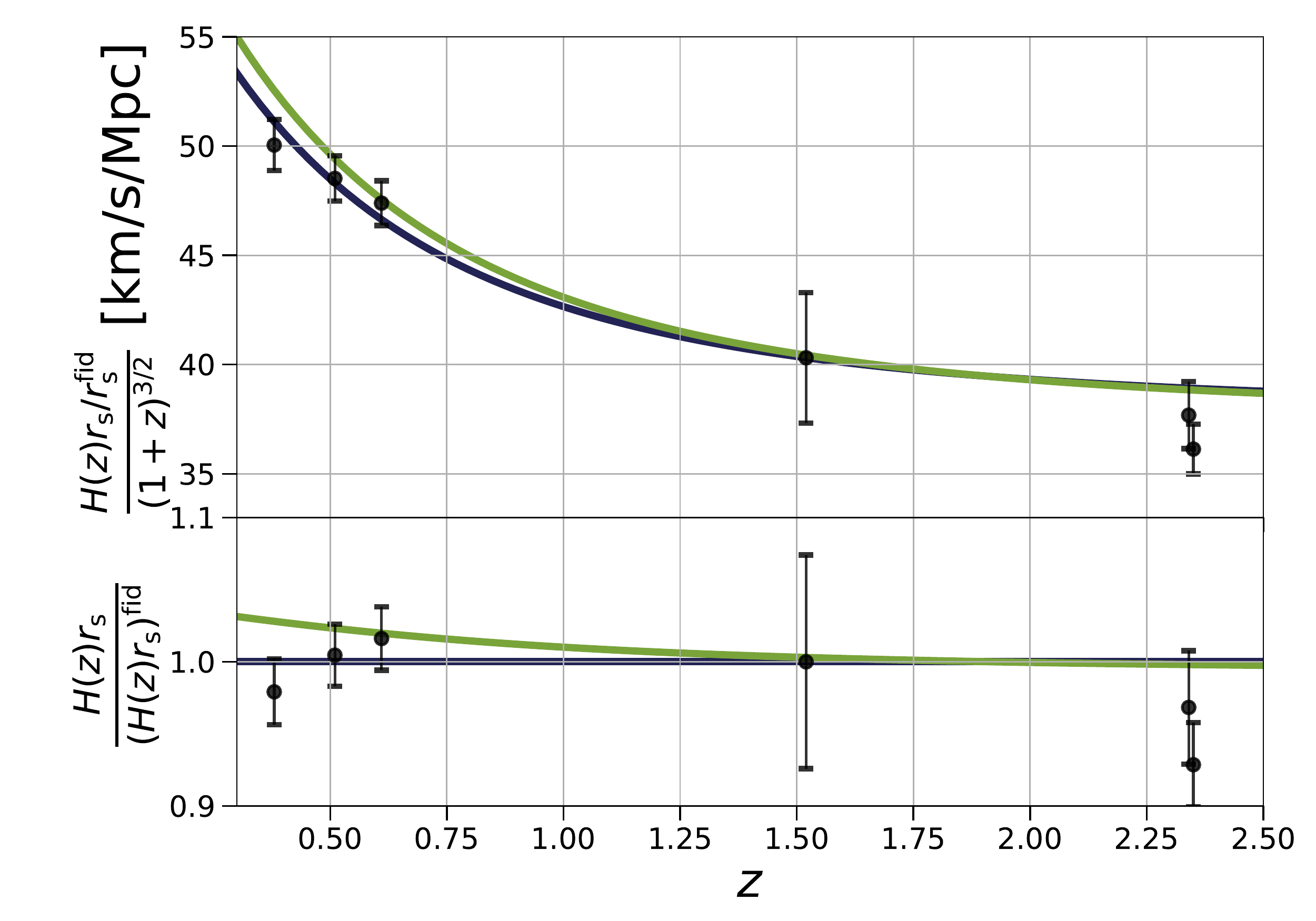}
\includegraphics[width=\columnwidth]{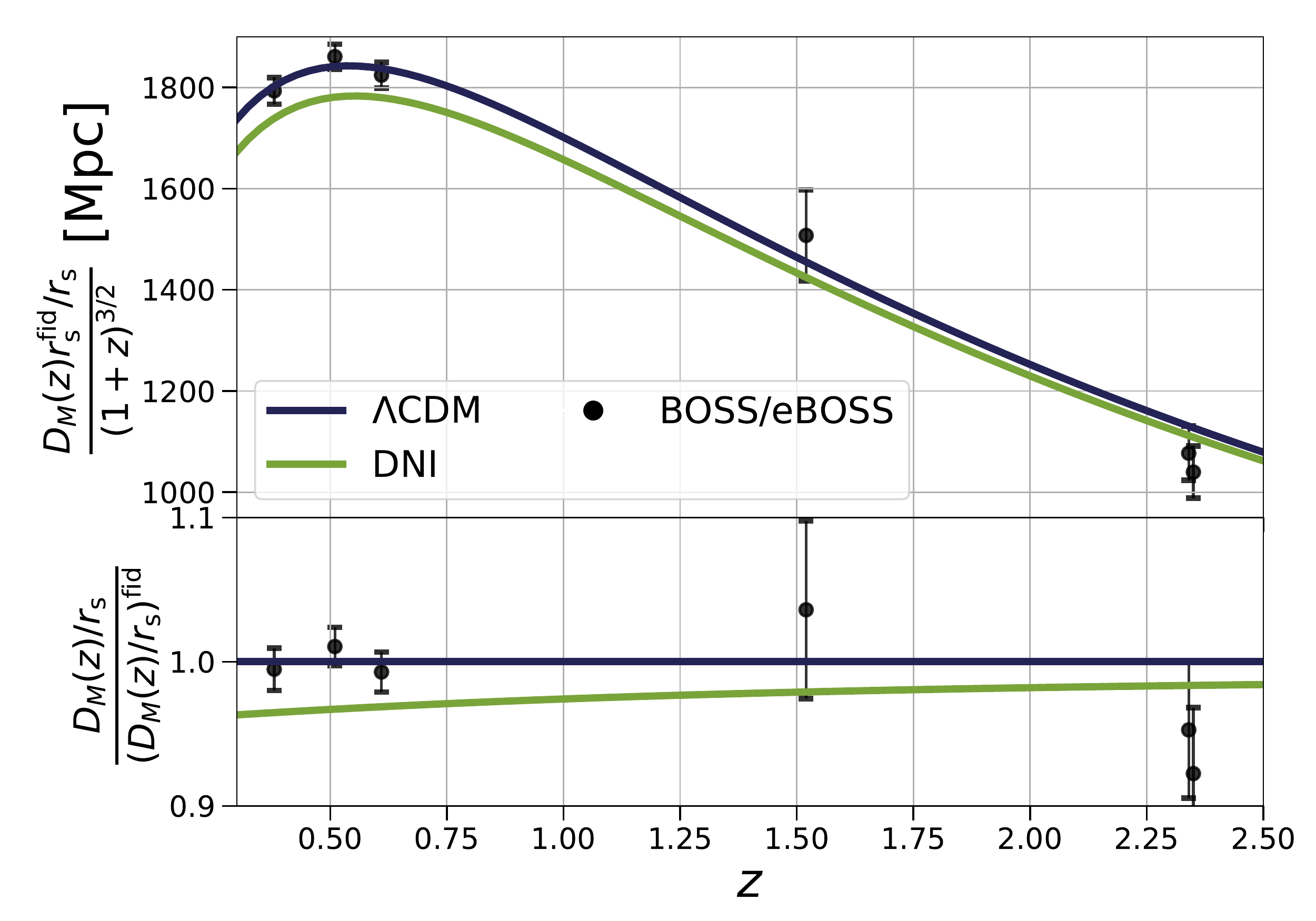}
\caption{Hubble expansion rate over $(1+z)^{3/2}$(top) and comoving angular diameter distance over $(1+z)^{3/2}$ (bottom) as a function of redshift, weighted by the ratio between the actual sound horizon at radiation drag and its fiducial value ($\Lambda$CDM best fit to \textit{Planck}). We show predictions for $\Lambda$CDM (blue) and DNI (green), as well as existing measurements from BOSS~\cite{Alam_bossdr12} and eBOSS\cite{GilMarin_QSOeBOSS,eBOSS_Lyalpha_auto,eBOSS_Lyalpha_cross}. Note that the error bars shown here do not include the covariance between measurements.}
\label{fig:DAHz}
\end{figure}

Our findings ease the worry regarding the validity of reported BAO measurements assuming a $\Lambda$CDM template for models beyond $\Lambda$CDM with large deviations at the perturbation level. This concern has prompted to some studies to omit BAO measurements from the set of cosmological data used to constrain cosmological models (see e.g., Ref.~\cite{Ghosh_DNIH0}). We argue that  BAO measurements can and should be included, especially given their importance in constraining proposed solutions to the $H_0$ problem~\cite{BernalH0,Aylor_sounds,Know_H0hunter}.

In particular, let us consider how the best-fit DNI model obtained in Ref.~\cite{Ghosh_DNIH0} without including BAO measurements fits the BAO observations. In Fig.~\ref{fig:DAHz}, we show the expansion history as predicted by the best-fit $\Lambda$CDM cosmology to \textit{Planck} 2018 and the best-fit DNI model to data from \textit{Planck} 2015~\cite{Planckparameterspaper}, WiggleZ dark energy survey~\cite{Parkinson_wigglez}, and SH0ES~\cite{RiessH0_19} (the data set considered in Ref.~\cite{Ghosh_DNIH0}). We also show existing measurements from BOSS~\cite{Alam_bossdr12} and eBOSS~\cite{GilMarin_QSOeBOSS,eBOSS_Lyalpha_auto,eBOSS_Lyalpha_cross}, demonstrating that $\Lambda$CDM provides a much better fit. The DNI prediction implies a $\chi^2 = 5.9$ larger than the $\Lambda$CDM prediction for the combined likelihood of BOSS~\cite{Alam_bossdr12}, 2dFGS~\cite{Beutler11} and SDSS DR7 MGS~\cite{Ross15} (the BAO data set considered in \textit{Planck} analyses). We conclude that BAO measurements disfavor the reported best-fit DNI model. We expect that BAO measurements favor DNI models with lower $H_0$ than the reported best fit (and corresponding lower $u_{\rm DNI}$ and $f_{\rm DNI}$), closer to the $\Lambda$CDM constraint.

This expectation is common to any model that modifies pre-recombination physics without altering the $\Lambda$CDM prediction of $r_{\rm d}$. This is due to the strong, model-independent constraint set by BAO on $r_{\rm d}h$~\cite{StandardQuantities}. This result hints that only introducing a phase shift in the acoustic perturbations without modifying $r_{\rm d}$ is not enough, and that a $r_{\rm d}$ needs to be lower  to solve the $H_0$ tension~\cite{BernalH0,Poulin_H0,Aylor_sounds,Know_H0hunter}.

We note however that cosmological parameter inference could be systematically affected by coherent small biases in the determination of $\alpha_\perp$ and $\alpha_\parallel$ at several redshifts.  We explore this possibility adapting the estimation of the systematic bias in parameter inference due to inaccurate approximations presented in Ref.~\cite{Bernal_Fisher} to our needs. The systematic shift in the cosmological parameters $\bm{\vartheta}$ when the measured rescaling parameters $\bm{\alpha}^{\rm m}$ are biased with respect to their true values $\bm{\alpha}^{\rm tr}$ can be estimated as 
\begin{equation}
\begin{split}
  \Delta\bm{\vartheta}= & \left(\sum_SF_S\right)^{-1}\times \\
& \times\sum_{S,i,j}\left(\nabla_\vartheta\alpha_{S;i}\right)\left({\rm Cov_S}\right)^{-1}_{ij}\left(\alpha^{\rm m}_j-\alpha^{\rm tr}_{S;j}\right),
\end{split}
\label{eq:estimbias}
\end{equation}
where $\Delta\bm{\vartheta}=\bm{\vartheta}^{\rm m} - \bm{\vartheta}^{\rm tr}$ (with $\bm{\vartheta}^{\rm m}$ denoting the parameters inferred from the observations without accounting for the bias in the analysis), $F$ is the Fisher matrix of the cosmological parameters, Cov is the covariance matrix of the measurements of the rescaling parameters, $i$ and $j$ are the entries of this covariance and the $\bm{\alpha}$ vectors, and $S$ refers to each of the surveys or independent measurements included in the analysis. 

Consider the DNI cosmologies from Table~\ref{tab:models},  with the measurements and biases reported in Fig.~\ref{fig:alphas}. In this case, we have four independent measurements and two relevant cosmological parameters: the total matter density parameter $\Omega_{\rm M}$ today and $H_0$. We obtain each Fisher matrix transforming the corresponding $\left({\rm Cov_S}\right)^{-1}$ to the $\Omega_{\rm M}$ - $H_0$ space. We find biases of $\sim -0.1\sigma$ ($\sim 0.1 \sigma$) and $\sim -1.4\sigma$ ($\sim -0.3\sigma$) for $\Omega_{\rm M}$ and $H_0$, respectively, for the DNI 2 (1) cosmology. When only the BOSS surveys are considered, these biases reduce to $\sim -0.01\sigma$ ($\sim -0.03\sigma$) and $\sim -0.2\sigma$ ($\sim -0.01\sigma$). All the biases are compared with the corresponding uncertainties from the cosmological parameter inference.

Here, we consider the bias obtained when using only the BAO measurements from the surveys used in this work; the effective bias affecting $\bm\vartheta$ when combining additional cosmological probes can be estimated in a similar way following Ref.~\cite{Bernal_Fisher}. Nonetheless, when different probes are combined, small systematic errors may lead to significant biases in the joint parameter inference. The study performed here to discuss potential biases on the cosmological parameters can be adapted to estimate the associated potential systematic error budget sourced by employing standard BAO measurements to constrain models beyond $\Lambda$CDM. This error budget may vary from case to case, depending on the cosmological model under study and the actual observations used, and might be important for the analysis of future BAO measurements.

\subsection{Amending biases on  standard BAO analyses}
If there is a situation in which a significant bias on $\alpha_\perp$ or $\alpha_\parallel$ is indeed found when applying the analysis presented in this work, the power spectrum must be reanalyzed to measure the BAO scale. There are a few different options about how to proceed. 

One option is to find a template assuming a different fiducial cosmology that  minimizes the bias seen in the analysis and refit the BAO with this template. In this case, $\alpha_\perp$ and $\alpha_\parallel$ would be given by Eq.~\eqref{eq:scaling_2cosmo}. This new fiducial cosmology will be most likely beyond $\Lambda$CDM. 
However, the impact of the feature not captured by a rescaling may depend on cosmological parameters. In this case, a fixed template without modeling such a feature would still fail to remove the bias in $\alpha_\perp$ and $\alpha_\parallel$.

Hence, a second option consists of adding freedom to the analysis. In case the extra contribution to $O_{\rm lin}$ can be robustly modeled, it is possible to include it in the BAO analysis with one or more new nuisance parameters that control its impact. Using this approach, however, would weaken the constraints on $\alpha_\perp$ and $\alpha_\parallel$, given the addition of extra nuisance parameters. 

Finally, if a reliable model describing the extra feature cannot be found, the analysis can be performed changing the template at each step of the MCMC. This procedure is related to alternative  methodologies to extract cosmological information from the BAO feature that do not rely on a pre-computed template (see e.g., Refs.~\cite{Damico_EFTBAO,Ivanov_EFTBOSS}). These methods do not aim to obtain a model-independent measurement of the expansion history of the Universe, nor can they extract agnostic independent information of early-time physics through $r_{\rm d}$, since the template readjusts during the fitting procedure. In these cases, a prior is needed to break the degeneracy between at least $\Omega_{\rm cdm}h^2$ and $\Omega_{\rm b}h^2$. This prior may either come from CMB observations and be directly applied to $r_{\rm d}$, or be obtained  from primordial deuterium and helium abundances assuming standard Big Bang nucleosynthesis and be applied to $\Omega_{\rm b}h^2$ and $N_{\rm eff}$.

\subsection{BAO cosmology beyond $\pmb{\alpha_\perp}$ and $\pmb{\alpha_\parallel}$}

Even if there is no bias in the BAO measurements, there can still be cosmological signatures that are not captured by the modeling of the rescaling of the template as described in Section~\ref{sec:analysis}. This is the case for $\Lambda$CDM+$N_{\rm eff}$, for instance: varying $N_{\rm eff}$ changes the phase shift of the BAO in a way that cannot be reproduced by a rescaling with $\alpha_\perp$ or $\alpha_\parallel$. Additional freedom is needed in order to consistently include this effect in the BAO analysis. This was explored in Ref.~\cite{Baumann_NeffBAO_fisher}, where an extra parameter was included as a multiplicative factor of a $k$-dependent function that modeled the amount of phase shift. 

Once this additional freedom is included in the analysis, instead of considering the extra parameter as a nuisance and marginalizing over it, it can also be treated as the parameter of interest. In this case, one would marginalize over the rest of the parameters to constrain the amplitude of the extra feature introduced in $O_{\rm lin}$. This is the case for the phase shift induced by $N_{\rm eff}$ in Ref.~\cite{Baumann_NeffBAO_measure}, where a phase shift was detected at 95\% confidence level after applying CMB priors on $\alpha$. Since most of these effects are expected to be isotropic (affecting only $O_{\rm lin}$ and not the Alcock-Paczynski effect or the redshift-space distortions), it may be preferable to perform an isotropic BAO analysis in order to reduce degeneracies between parameters.

\section{Conclusions}\label{sec:conclusions}
BAO provide a robust probe of the expansion history of the Universe at low redshift, as well as a calibrator of the sound horizon at radiation drag.  Recently, the importance of BAO measurements has been highlighted again with a crucial role in the resolution of the $H_0$ tension, which can be reframed as a mismatch between the two anchors of the cosmic distance ladder: $H_0$ and $r_{\rm d}$~\cite{BernalH0}. These are precisely the two quantities which set the normalization of the expansion history of the Universe as constrained by BAO measurements.

Standard BAO analyses are robust and model independent, since they rely on a template fitting using rescaling parameters, which encode the cosmological information of interest. In addition, differences in the shape between the measured power spectrum and the  pre-computed template (sourced by the choice of a wrong fiducial cosmology or uncertainties in small-scale clustering) are   marginalized over using a plethora of nuisance parameters. As extensions of $\Lambda$CDM become more complex in the search of new physics or in the attempt to resolve cosmological tensions,  this approach might not be flexible enough to cover these cosmologies, rendering BAO measurement unsuitable for studying such models. 

In this work, we have assessed the flexibility of the standard BAO analysis under different cosmological models: $\Lambda$CDM, $\Lambda$CDM+$N_{\rm eff}$, EDE, and DNI. We have performed mock analyses of galaxy power spectra, using the same $\Lambda$CDM template for all of them. We find that, for the models explored in this work, there is no bias in the BAO analyses for existing data. Moreover, the maximum bias found for future surveys is $\sim 1\sigma$ for an extreme DNI model which provides a very bad fit  to \textit{Planck} data. Therefore, our findings reinforce the model independence and robustness of the standard BAO analysis, and they support using reported BAO measurements in terms of $\alpha_\perp$ and $\alpha_\parallel$ (obtained using a $\Lambda$CDM template) to constrain these models. This will be further supported by the implementation of blinding analyses in future galaxy surveys~\cite{Brieden_blinding}.

The models considered in this work have been chosen as a representative sample of models that aim to solve the $H_0$ tension through modifications prior to recombination. Nonetheless, our results can be extrapolated to other models that do not impact the clustering of dark matter and baryons with new physics after recombination. Moreover, futuristic surveys may reach a level of precision for which the low significance of the biases found in this work is enough to affect the measurement.  If there is a model with a modified perturbation history that changes the BAO feature beyond rescaling factors, we advocate for the methodology described in this work to ascertain whether reported constraints in the $\alpha_\perp$-$\alpha_\parallel$ plane are valid for cosmological parameter inference. In case a bias in the BAO fit is found, we have suggested possible ways forward to add flexibility to the modeling of the power spectrum and perform an unbiased BAO measurement. 

Next-generation galaxy surveys like DESI~\cite{desi}, Euclid~\cite{euclid} and SKA~\cite{redbook} will improve upon existing BAO measurements through higher precision observations and reaching higher redshifts. Additionally, future line-intensity mapping experiments promise to extend BAO measurements up to $z\sim 9$ with competitive precision~\cite{Bernal_IM,Bernal_IM_letter}.
 Checking the validity of BAO analyses for exotic models is of the utmost importance in order to exploit the promising outlook for measurements in the near future, especially given the importance of BAO as a cosmological probe.

\acknowledgments
We thank Licia Verde, H\'{e}ctor Gil-Mar\'{i}n and Vivian Poulin  for useful discussions, and Graeme Addison and Alvise Raccanelli for comments on the last stages of the manuscript. JLB is supported by the Allan C. and Dorothy H. Davis Fellowship. TLS is supported by the Research Corporation and NASA ATP grant number 80NSSC18K0728.

\appendix
\section{Extraction of the smooth power spectrum $\pmb{P_{\rm m, sm}}$}\label{sec:Olin_templ}
The idea behind the extraction of the smooth matter power spectrum $P_{\rm m,sm}$ from the total matter power spectrum $P_{\rm m}$ is conceptually simple: it is based on removing the wiggles produced by the BAO. However, rather than performing a brute-force smoothing of $P_{\rm m}$ or interpolating using only the zero points between the wiggles, one can choose more stable and efficient options. There are several procedures to extract $P_{\rm m,sm}$, mainly divided in two broad groups. One option relies on the computation of the matter power spectrum without including the BAO contribution, which can be done, e.g., using the analytic transfer function derived in Ref.~\cite{EH98}. Alternatively, one can use Eq.~\eqref{eq:pktoxi} to convert $P_{\rm m}$ into $\xi_{\rm m}(r)$, directly remove the BAO peak from the correlation function, and transform back to $P_{\rm m}$. We adopt the latter method in this work. We refer the interested reader to Ref.~\cite{Kirkby_BAOtemplate} for a detailed comparison between different methodologies.

\begin{figure}[ht!]
\centering
\includegraphics[width=\columnwidth]{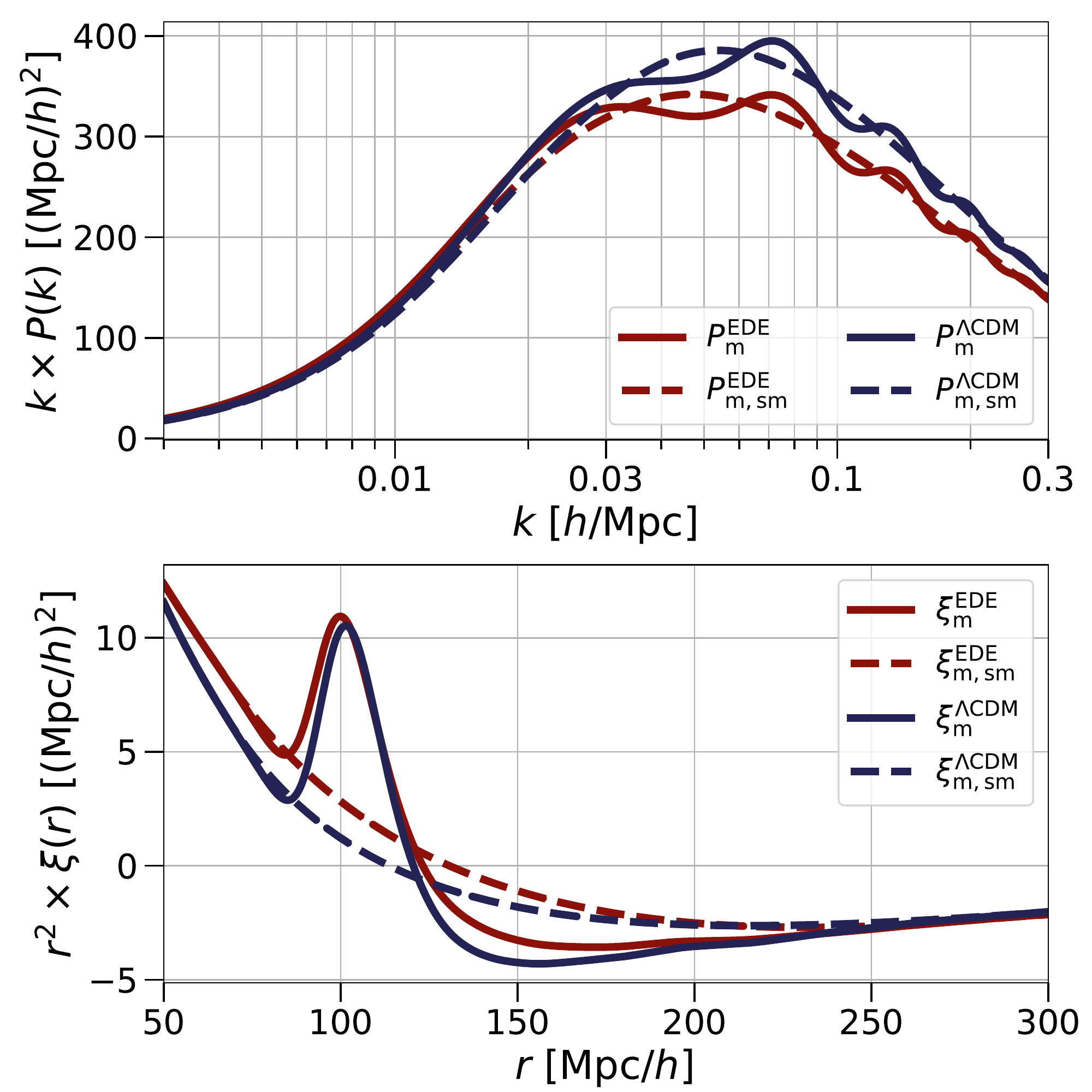}
\caption{Top: Total linear matter power spectrum before (solid lines) and after (dashed lines) the removal of the BAO contribution ($P_{\rm m}$ and $P_{\rm m,sm}$, respectively). Bottom: Same as top but in configuration space. Factors of $k$ and $r^2$ are added respectively for clarity. Blue lines correspond to $\Lambda$CDM predictions and red lines to EDE predictions (with $f_{\rm EDE}=0.3$).}
\label{fig:BAO_removal}
\end{figure}

Since we are interested in exploring cosmologies very different from $\Lambda$CDM, especially regarding the BAO scale and feature, we avoid fits to polynomials at fixed scales, as done in, e.g., Ref.~\cite{Kirkby_BAOtemplate}. Instead, we use a more flexible methodology described below:
\begin{enumerate}
\item Obtain the total matter power spectrum $P_{\rm m}$ from a Boltzmann solver and use Eq.~\eqref{eq:pktoxi} to obtain the corresponding correlation function, $\xi_{\rm m}$. We use the public \texttt{mcfit} python package to evaluate integrals, such as the one appearing in Eq.~\eqref{eq:pktoxi}.\footnote{\url{https://github.com/eelregit/mcfit}.}  In order to avoid numerical noise, $P_{\rm m}$ needs to be evaluated at a large number of points $N$ and for a wide range in $k$. We sample $k$ uniformly in logspace in $N=4096$ points in the interval $k\in \left[10^{-5},20\right]$~$h/$Mpc.
\item Take the corresponding entries to $r_{1}\in\left[60,70\right]$ Mpc/$h$ and $r_2\in\left[200,300\right]$ Mpc/$h$ in the $r$ array, $i_{1}$ and $i_2$, respectively, and interpolate $r^2\xi_{\rm m}$ evaluated in the interval $\left[i=0,i_1\right]\bigcup\left[i_2,i=N-1\right]$ using  cubic splines. 
\item Evaluate the interpolation object in the original $r$ array and remove the $r^2$ factor, so that a smooth correlation function $\xi_{\rm m,sm}$ without the BAO peak is obtained.
\item Transform $\xi_{\rm m,sm}$ back to Fourier space, i.e., obtaining $P_{\rm m,sm}$. $O_{\rm lin}$ is obtained by computing the ratio between the total matter power spectrum $P_{\rm m}$ and $P_{\rm m,sm}$.
\item Iterate points 2-4 varying $r_1$ and $r_2$ to optimize the convergence of $O_{\rm lin}=1$ at $k\lesssim 10^{-3}$ $h$/Mpc: at scales much larger than $r_{\rm d}$ there is no effect from the BAO and the smooth and actual clustering must be equal.
\end{enumerate}
 We compare the total and the smooth matter power spectrum and correlation function in Fig.~\ref{fig:BAO_removal}. Note that $r_1$ is smaller than the scale of the local minimum of $\xi_{\rm m}$, and the range of $r_2$ is significantly larger than $r_{\rm d}$. This is because the two minima of $\xi_{\rm m}$ are deeper than for $\xi_{\rm m,sm}$ due to the enhanced clustering around $r_{\rm d}$.

The computation of the smooth power spectrum or correlation function may be critical for the analysis of real observations, and the performance of different procedures must be compared. We have checked that our results are robust to the choice of the algorithm used to extract the broadband of the clustering, since the mock power spectrum is computed with the same modeling that is used in the likelihood.

\section{Mock $\pmb{P_{\rm g}^{\rm data}}$ computation}
\label{sec:mockdata}
As explained in Section~\ref{sec:bias_alphas}, we perform a mock BAO analysis in order to test for the presence of a bias in the measurement of the rescaling parameters $\alpha_\perp$ and $\alpha_\parallel$ when the measured power spectrum does not correspond to the fiducial cosmology assumed for the computation of the template in Eq.~\eqref{eq:Pkmu}. To do so, we need to compute a measured power spectrum for different cosmologies. In this Appendix we explain our procedure to compute this mock $P_{\rm g}^{\rm data}$.

First, we compute $P(k,\mu)$ following Eq.~\eqref{eq:Pkmu} with the cosmological and nuisance parameters corresponding to the target cosmological model we want to mock. However, we need to account for the fact that this power spectrum corresponds to a cosmology different from the assumed fiducial cosmology used to measure it. This means that the mock  observed $P(k,\mu)^{\rm data}$ is affected by the Alcock-Paczynski effect and the isotropic dilation. Note that in this case, $P(k,\mu)$ has been obtained with its corresponding $r_{\rm d}$ (instead of the fiducial template), so that we rescale $k$ with $q_\perp$ and $q_\parallel$ from Eq.~\eqref{eq:scaling_q}. With this correction, we compute $\bm{P}_{\rm g}^{\rm data}$ using Eq.~\eqref{eq:multipole_scale}.  Moreover, given that the power spectrum is measured in $\left({\rm Mpc}/h\right)^3$, for the fiducial value of $h$, we need to rescale the units accordingly. Finally, we assume the covariance would be computed from mock galaxy catalogs; hence, it would correspond to the fiducial cosmology. Therefore, we compute the covariance by applying Eq.~\eqref{eq:covariance} to the fiducial power spectrum. 

In order to evaluate the bias without any confusion introduced by statistical dispersion around the true values, we decide not to include dispersion on the mock power spectrum around the predicted theoretical power spectrum. Therefore, even though we consider the correct covariance, all the points of the mock power spectrum coincide with its theoretical prediction. Nonetheless, we describe here how to include the statistical dispersion of the mock power spectrum around the theoretical prediction, for completeness. 
Note that the covariance of the power spectrum is not diagonal (Eq.~\eqref{eq:covariance}), so we need to take into account the covariance between multipoles in order to mock the dispersion about the true values of $\bm{P}_{\rm g}^{\rm data}$. Therefore, we diagonalize the covariance matrix and draw random values from Gaussian distributions centered at 0 and with variances corresponding to each of the entries of the diagonalized covariance. Then, we transform this vector back to the original $k$ base and add it to $\bm{P}_{\rm g}^{\rm data}$. The resulting dispersion around $\bm{P}_{\rm g}^{\rm data}$ properly accounts for the covariance between the multipoles.

\section{Connection between CMB and BAO}
\label{sec:CMBtoBAO}

Under $\Lambda$CDM, the BAO feature present in the matter power spectrum at low redshift comes from the same dynamics that produce the acoustic structure measured in the CMB: baryons and photons are tightly coupled in the early Universe and undergo acoustic oscillations, which become imprinted in both the matter power spectrum and the photon temperature fluctuations of the CMB. In fact, the use of a $\Lambda$CDM template when analyzing the BAO feature relies on the assumption that if $\Lambda$CDM provides a good fit to the CMB power spectra, it also provides a good fit to the BAO feature. With this qualitative fact in mind, it is useful to quantify this relationship in light of the analysis we have presented in this work.

During radiation domination, and after horizon crossing, the density contrast $\delta_{\rm c}$ of cold dark matter starts to grow logarithmically with the scale-factor, $\delta_{\rm c} \propto \ln(a)$. As time passes, smaller-scales modes start collapsing as $\delta_{\rm c} \propto a$ so that, after matter-radiation equality, all modes within the horizon grow as $\delta_{\rm c} \propto a$.  Baryons, on the contrary, only fall into the dark matter potential wells after they have recombined and got released from the radiation pressure. 
  Then, after decoupling, the evolution of cold dark matter and baryons  is given by \cite{Meszaros:1974tb,Ma:1995ey} 
\begin{eqnarray}
\ddot \delta_{\rm c} + \frac{\dot a}{a} \dot \delta_{\rm c} & = \frac{3}{2} \left(\frac{\dot a}{a}\right)^2 \left[R_{\rm c} \delta_{\rm c}+(1-R_{\rm c}) \delta_{\rm b}\right],\label{eq:Meszaros1}\\
\ddot \delta_{\rm b} + \frac{\dot a}{a} \dot \delta_{\rm b} & = \frac{3}{2} \left(\frac{\dot a}{a}\right)^2 \left[R_{\rm c} \delta_{\rm c}+(1-R_{\rm c}) \delta_{\rm b}\right],
\label{eq:Meszaros2}
\end{eqnarray}
where  $R_{\rm c} \equiv \rho_{\rm c}/\rho_{\rm m}$ is the fraction of the cold dark matter energy density $\rho_{\rm c}$ to the total matter density $\rho_{\rm m} = \rho_{\rm c} + \rho_{\rm b}$ of cold dark matter and baryons, and the dot denotes a derivative with respect to conformal time, $\eta$. Note that at decoupling the universe is matter dominated so that $\dot a/a = 2/\eta$.

The total matter transfer function is given by the weighted sum of the cold dark matter and baryon transfer functions:
\begin{equation}
T_{\rm m} = R_{\rm c} \delta_{\rm c} + (1-R_{\rm c}) \delta_{\rm b},
\end{equation}
where $\delta_{\rm c}$ and $\delta_{\rm b}$ are given by the solution of the coupled Eqs.~\eqref{eq:Meszaros1} and \eqref{eq:Meszaros2}. If we assume tight-coupling between baryons and photons, adiabatic initial conditions and instantaneous decoupling,  so that $\delta_{\rm b}$ is related with the radiation overdensity $\delta_\gamma$ as $\delta_{\rm b,dec} = 3\delta_{\gamma,{\rm dec}}/4$ (where the subscript `dec' refers to quantities evaluated at decoupling), we obtain
\begin{equation}
T_{\rm m} = \mathcal{A}\left(\frac{\eta}{\eta_{\rm dec}}\right)^2 ,
\end{equation}
 and  
\begin{equation}
\begin{split}
\mathcal{A} \equiv & \frac{1}{20}\left[4R_{\rm c} \left(3 \delta_{\rm c,dec}+\eta_{\rm dec}\dot{\delta}_{\rm c,dec}\right)\right. +\\ 
&\left.+3\left(1-R_{\rm c}\right)\left(3\delta_{\gamma,{\rm dec}}+\eta_{\rm dec} \dot{\delta}_{\gamma,{\rm dec}}\right)\right],
\end{split} 
\end{equation}
where we have neglected decaying solutions (assuming $\eta\gg\eta_{\rm dec}$) and the effect of the cosmological constant. The linear matter power spectrum thus takes the form 
\begin{equation}
P_{\rm m}(k) = P_{\rm prim}(k) T_{\rm m}^2(k,\eta) = P_{\rm m,sm}(k)O_{\rm lin}(k),
\end{equation}
where  $P_{\rm prim}(k) = 2\pi^2A_s k^{-3} (k/k_p)^{n_s -1}$ is the primordial power spectrum (where $k_p$ is the pivot wave number, typically taken to be $k_p = 0.05\ {\rm Mpc}^{-1}$) and
\begin{eqnarray}
\frac{P_{\rm m,sm}}{P_{\rm prim}} &=&\frac{R_{\rm c}^2}{25} \left(3 \delta_{\rm c,dec} + \eta_{\rm dec} \dot{\delta}_{\rm c,dec}\right)^2 \left(\frac{\eta}{\eta_{\rm dec}}\right)^4,\\
O_{\rm lin}-1&\simeq & \frac{3(1-R_{\rm c})( 3\delta_{\gamma,{\rm dec}} + \eta_{\rm dec} \dot{\delta}_{\gamma,{\rm dec}})}{2 R_{\rm c} (3\delta_{\rm c,dec}+\eta_{\rm dec} \dot{\delta}_{\rm c,dec})}\,, \label{eq:Olin}
\end{eqnarray}
where we have neglected the term  proportional to $(1-R_{\rm c})^2$. 
Let us now focus on the BAO feature, whose scale dependence is given by $\delta_{\gamma,{\rm dec}}$ and $\dot{\delta}_{\rm \gamma,dec}$. Note also that this approximate $O_{\rm lin}$  does not evolve with time, as discussed for the numerically obtained  $O_{\rm lin}$ described in Appendix~\ref{sec:Olin_templ} and shown in Fig.~\ref{fig:Olin_rescaled}. During tight-coupling, the acoustic oscillations in the photon-baryon fluid  are damped along with a scale-dependent enhancement due to a driving force: 
\cite{Hu:1996mn,Hu:2000ti,Dodelson:2003ft}
\begin{eqnarray}
\delta_{\gamma,{\rm dec}} &\simeq& -\mathcal{D}(k) \cos(c_{\rm s} k \eta_{\rm dec}),\label{eq:dgamma}\\
\dot{\delta}_{\gamma,{\rm dec}} &\simeq& \mathcal{D}(k) c_{\rm s} k \sin(c_{\rm s} k \eta_{\rm dec}),\label{eq:dotdgamma}
\end{eqnarray}
where $c_{\rm s}$ is the photon-baryon sound speed, and $\mathcal{D}$ encodes the Silk damping and the scale-dependent enhancement.  Note that the sound horizon at decoupling is $r_{\rm s} = c_{\rm s}  \eta_{\rm dec}$.

The acoustic features in the CMB arise from different, but related, terms. If we assume again an instantaneous decoupling, the visibility function is a Dirac delta function: $g(\eta) = \delta_D(\eta-\eta_{\rm dec})$. Under this approximation, the CMB power spectra are given by~\cite{Seljak:1994yz,Zaldarriaga:1995gi,Seljak:1996is} 
\begin{equation}
C_\ell^{XY}  = (4\pi)^2  \int k^2 dk P_{\rm prim}(k) \Delta_{\ell}^{X}(k) \Delta_{\ell}^{Y}(k) 
\label{eq:CMBCls}
\end{equation}
with
\begin{align}
\Delta_\ell^T(k) \approx & \left[\psi_{\rm dec} + \frac{1}{4}\delta_{\gamma,{\rm dec}}\right] j_\ell(x_0) +v_{\rm b,dec} j^\prime_\ell( x_0),\\
\Delta_\ell^E(k) \approx & \sqrt{\frac{(\ell+2)!}{(\ell-2)!}}  \frac{4}{3} \Pi_{\rm dec}\frac{j_\ell(x_0)}{x_0^2},
\end{align}
where ${\psi}$ is the Newtonian potential transfer function, $v_{\rm b}$ is the baryon velocity perturbation transfer function, ${\Pi}$ is the polarization transfer function, $x_0 \equiv k(\eta_0 - \eta_{\rm dec})$ (where $\eta_0$ is the conformal time today), and $^\prime$ denotes the derivative with respect to $x_0$.  
To leading order in $\dot \tau/(aH)$ we have 
\begin{eqnarray}
v_{\rm b,dec} &\simeq& \frac{3}{4} \frac{\dot \delta_{\gamma,{\rm dec}}}{k}\\
{\Pi}_{\rm dec} &\simeq& \frac{3}{4}\frac{\dot \delta_{\gamma,{\rm dec}}/4}{\dot \tau}, 
\end{eqnarray}
where $\dot \tau$ is the differential optical depth for Thomson scattering.

Combining Eq.~\eqref{eq:Olin} with Eqs.~\eqref{eq:dgamma} and~\eqref{eq:dotdgamma}, we can see that the ratio between the terms in the second bracket, $\lvert \eta_{\rm dec}\dot{\delta}_{\gamma, {\rm dec}}/3\delta_{\gamma,{\rm dec}}\lvert = r_{\rm s}k/3\simeq k/(0.02\, {\rm Mpc^{-1}})$. This shows that, in the scales of interest for the acoustic feature, $O_{\rm lin}$ is driven by the term depending on $\dot{\delta}_{\rm\gamma,dec}$. Therefore, comparing with Eq.~\eqref{eq:CMBCls}, the acoustic oscillations in $O_{\rm lin}$ are a factor $\pi/2$ out of phase with respect to the CMB temperature anisotropies,  which is a well-known result (see, e.g., Ref.~\cite{Eisenstein_BAOrobust}). 

We can see that the CMB and BAO acoustic features are, to first approximation, given by the photon density perturbation and its first time derivative at decoupling. Therefore, within the context of the basic approximations made here, we can expect that any cosmological model, even beyond $\Lambda$CDM,  providing a good fit to the CMB spectra should also provide a good fit to the BAO feature. This argument also supports the extrapolation of the conclusions of this work beyond the four cosmological models considered.

Nonetheless, there are some differences to note between the acoustic features in the BAO and the CMB. 
One difference is that the CMB spectrum is a  projected version, and the Fourier modes are integrated over spherical Bessel functions. Even though $\ell \simeq k(\eta_0 - \eta_{\rm dec})$, wave numbers in the range $\Delta k \simeq k/2$ contribute to a measured $\ell$. A more dramatic difference is related to the amplitude of the oscillations: while CMB oscillations are at least order $\sim 1$,  BAO oscillations have an amplitude $\sim 0.1$ (see Fig.~\ref{fig:Olin_rescaled}). In addition to this, the nuisance parameters necessary to robustly extract information from the BAO feature make current measurements of the BAO less sensitive to changes  (beyond a rescaling of distances with $r_{\rm d,s}$)  in the acoustic oscillations than the CMB.  This is why, even for cosmic variance limited BAO measurements over large volumes and wide redshifts ranges, CMB priors related with the background expansion are needed in order to obtain competitive cosmological constraints from the shape of the BAO (see e.g.,~\cite{Baumann_NeffBAO_fisher} for the case of $N_{\rm eff}$). 

\bibliography{biblio}
\bibliographystyle{utcaps}

\end{document}